\documentclass[preprint,11pt]{elsarticle}
\usepackage{amsmath}
\usepackage{amssymb}
\usepackage{lineno}
\usepackage{hyperref}
\usepackage{color,soul}
\usepackage{textcomp} 
\usepackage{acronym}
\usepackage[labelsep=period]{caption}
\usepackage{color}
\usepackage[inline]{enumitem}
\usepackage{float}
\usepackage{epstopdf}
\usepackage{graphicx}
\usepackage{morefloats}
\usepackage[export]{adjustbox}
\usepackage{todonotes}
\usepackage{subfig}
\usepackage[linesnumbered,ruled,vlined]{algorithm2e}
\usepackage{textcomp}
\usepackage[autostyle]{csquotes}
\usepackage[many]{tcolorbox}
\newtcolorbox{highlighted}{colback=yellow,coltext=black,breakable}

\usepackage{cancel}
\usepackage{tikz}
\usetikzlibrary{matrix,calc}

\makeatletter
\newcommand{\mathleft}{\@fleqntrue\@mathmargin\parindent}
\newcommand{\mathcenter}{\@fleqnfalse}
\makeatother

\usepackage[margin=3.0cm]{geometry}

\setcitestyle{square}
\biboptions{numbers,sort&compress}
\modulolinenumbers[5]
\acrodef{NPR}{Negative Poisson's Ratio}
\acrodef{HPM}{Heaviside Projection Method}
\acrodef{NAND}{Nested ANalysis and Design}
\acrodef{SIMP}{Solid Isotropic Material with Penalization}
\acrodef{MMA}{Method of Moving Asymptotes}
\acrodef{GMRES}{Generalized Minimal RESidual}
\acrodef{BVP}{Boundary Value Problem}
\acrodef{RVE}{Representative Volume Element}
\acrodef{RUC}{Repeating Unit Cell}
\acrodef{DOF}{Degree-Of-Freedom}
\acrodef{PBC}{Periodic Boundary Condition}

\graphicspath{{./figs/}}

\makeatletter
\def\ps@pprintTitle{%
	\let\@oddhead\@empty
	\let\@evenhead\@empty
	\def\@oddfoot{\footnotesize\itshape
		{} \hfill\today}%
	\let\@evenfoot\@oddfoot
}
\makeatother

\begin{document}

\begin{frontmatter}

\journal{}
\title{Topology optimization of nonlinear periodically microstructured materials for tailored homogenized constitutive properties}

\author[UCSDMAE]{Reza Behrou}
\corref{mycorrespondingauthor}
\cortext[mycorrespondingauthor]{Corresponding author: Reza Behrou (rbehrou@eng.ucsd.edu)}
\author[UCSDMAE]{Maroun Abi Ghanem}
\author[UCSDMAE]{Brianna C. Macnider}
\author[UCSDMAE]{Vimarsh Verma}
\author[UCSDMAE]{Ryan Alvey}
\author[UCSDMAE]{Jinho Hong}
\author[UWME]{Ashley F. Emery}
\author[UCSDSE]{Hyunsun Alicia Kim}
\author[UCSDMAE]{Nicholas Boechler}

\address[UCSDMAE]{Department of Mechanical and Aerospace Engineering, University of California San Diego, La Jolla, CA, USA}
\address[UWME]{Department of Mechanical Engineering, University of Washington, Seattle, WA, USA}
\address[UCSDSE]{Department of Structural Engineering, University of California San Diego, La Jolla, CA, USA}

\begin{abstract}
A topology optimization method is presented for the design of periodic microstructured materials with prescribed homogenized nonlinear constitutive properties over finite strain ranges. The mechanical model assumes linear elastic isotropic materials, geometric nonlinearity at finite strain, and a quasi-static response. The optimization problem is solved by a nonlinear programming method and the sensitivities computed via the adjoint method. Two-dimensional structures identified using this optimization method are additively manufactured and their uniaxial tensile strain response compared with the numerically predicted behavior. The optimization approach herein enables the design and development of lattice-like materials with prescribed nonlinear effective properties, for use in myriad potential applications, ranging from stress wave and vibration mitigation to soft robotics. 
\end{abstract}


\end{frontmatter}

%
\section{Introduction}
\label{sec:Introduction}
The design of materials with tailored nonlinear properties is becoming increasingly important in materials sciences and engineering. 
This includes within the context of materials that exhibit constant properties over large deformation \cite{CWJ+:15}, novel wave tailoring behavior \cite{HLR:14}, multistability \cite{SKR+:15}, and the ability to match the nonlinear properties of biological media \cite{MCJ+:16}. Applications for these materials range from impact mitigation \cite{YMC+:19} to wearable electronics \cite{RHS:10}. One of the ways the realization of material nonlinearities has been achieved is through the introduction of a periodic microstructure, where the structure of the repeating unit cell experiences geometric nonlinearity under finite strain \cite{Nesterenko:13}. However, it remains challenging to identify, a priori, what specific structure is needed to obtain a specific, desired, effective nonlinear response. 

One method for designing materials with nonlinear responses is the use of topology optimization. Topology optimization is an affordable form-finding design methodology to obtain the optimized distribution of materials within a design domain \cite{BK:88}. Design of structures with geometric nonlinearity  via topology optimization has been successfully applied to a large class of structural problems such as stiffness design and compliant mechanisms \cite{BPS:00, BT:01, GL:01}. Extensions to the design of periodic microstructured materials with nonlinear responses have also been recently reported \cite{WSJ:14, MLR:13}. However, the examples of Refs. \cite{WSJ:14, MLR:13} were obtained for simplified physical responses such as design of periodic microstructures with tensile loading assumptions in \cite{WSJ:14}. 

An alternative approach is the use of nonlinear homogenization techniques that can be integrated into the formulation of the topology optimization problem. For the case of linear homogenization techniques, this approach has been successfully used for multiscale design optimization of structures with linear responses \cite{XB:17}. Under the assumption of finite strain theory, nonlinear homogenization techniques \cite{GKB:10, GKM+:17} have also been successfully integrated into topology optimization algorithms \cite{NTN:13, KWP+:19}. However, these methods involved simplifying assumptions. In Ref. \cite{NTN:13}, unit strains were assumed at the microstructure level, which mitigates the ability for a single structure to match tailored nonlinear load-displacement behavior. In Ref. \cite{KWP+:19}, the method was limited to longitudinal loading conditions.   

This paper presents a topology optimization method for the design of periodic microstructured materials for tailored nonlinear homogenized constitutive responses over finite strain ranges, where the effects of macroscopically varying local strains and/or stresses are considered on the response of the periodic unit cell. The strain/stress-driven nonlinear homogenization technique are considered, following the approach given by \cite{vanDijk:16}. The gradient-based topology optimization problem is formulated and the sensitivity equations are derived, allowing the design of nonlinear microstructured materials with tailored physical properties. The formulated topology optimization problem and the sensitivities are generalized for both material and geometric nonlinearities, considering the effects of macroscopically varying local strains and/or stresses on the response of \ac{RVE}. Altough the developed framework can be easily used for designing microstructures with material nonlinearity, e.g., hyperelastic and/or anisotropic materials, or design with imposed macroscale stresses, we only focus on the design of microstructured materials considering geometric nonlinearity under applied macroscale strains. The geometrically nonlinear behavior of microstructured \ac{RVE} is computed using total Lagrangian finite element formulation and the linearized forward finite element problem is solved by the arc-length method \cite{DCR+:12}. The effects of macroscale deformation and loading, i.e., applied macroscopic strain and stress, on the \ac{RVE} are considered through the \acp{PBC}. To alleviate numerical instabilities for an excessive deformation caused by low-density elements, a threshold on the elemental density, i.e., void region, is applied \cite{BRG:19}. The optimization problem is solved by a nonlinear programming method using the \ac{MMA} optimizer \cite{Svanberg:87}. To interpolate material properties, the \ac{SIMP} approach \cite{Bendsoe:89} is used. The projection method \cite{WLS:11} is used to regularize the optimization problem and converge to binary solutions (e.g., only one material is used in the structure, and each element is either material or void). Adjoint-based sensitivities are developed that derives the sensitivities of the nonlinear homogenized tangent stiffness tensor with respect to both state and design variables. Two-dimensional design examples and experimental calibrations are presented to study the performance of the presented approach and the response of topology optimized designs with tailored constitutive properties.
%

%
\section{Physical modeling}
\label{sec:NonlinearHomogenization}

\begin{figure}[t]
\centering
  \includegraphics[width=0.99\linewidth]{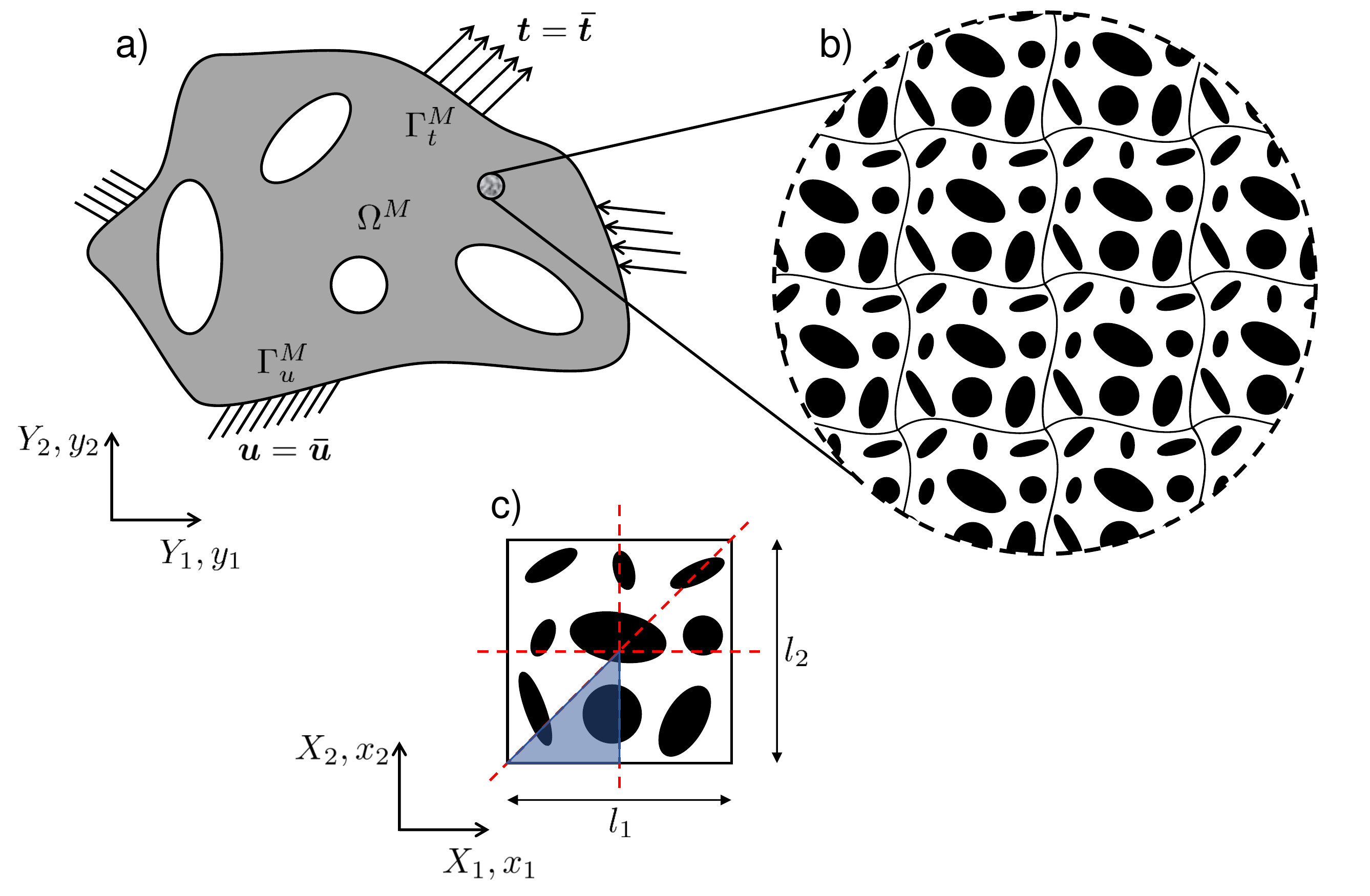}
\caption{a) Schematic representation of a 2D macroscale Boundary Value Problem (BVP) with a material point. $Y$ and $y$ denote the macroscale coordinates in reference and current configurations, $\boldsymbol{\bar{u}}$ and $\boldsymbol{\bar{t}}$ are the prescribed displacement and traction, applied on the macroscale boundaries $\Gamma_{u}^{M}$ and $\Gamma_{t}^{M}$, respectively, b) A periodically patterned microscale BVP at the material point, and c) Repeating Unit Cell (RUC) at the microscale. $X$ and $x$ denote the microscale coordinates reference and current configurations. The dashed lines represent the axes of symmetry and the shaded area is the design domain.}
\label{fig:01_MacroToMicroSchematic}
\end{figure}
\subsection{Macro- and microscale problems}
\label{sec:MacroMicroScaleProblem}
Consider macro- and microscale \acp{BVP} given in Fig. \ref{fig:01_MacroToMicroSchematic}. At both macro- and microscales, the deformation, in the absence of body forces and accelerations, is governed by the balance of linear momentum:
\begin{equation}\label{eq:MacroMicroScaleProblemEqs_01}
\begin{split}
\nabla_{0}^{k} \cdot \boldsymbol{P}^{k} = \boldsymbol{0},
\end{split}
\end{equation}
where $ k \in \left\{M, \mu\right\}$ represents either the macroscopic entities, denoted with $M$, or the microscopic ones, denoted with $\mu$, $\boldsymbol{P}$ is the first Piola-Kirchhoff stress tensor, and $\nabla_{0}$ is the gradient operator with respect to the reference configuration. For the macroscale \ac{BVP}, a constitutive relationship between stress and kinematic quantities can be developed through computational homogenization techniques that numerically extract the detailed computational response of a \ac{RVE} at the microscopic scale \cite{GKB:10, GKM+:17}. At any macroscopic material points, a periodically patterned \ac{RVE}, given in Fig. \ref{fig:01_MacroToMicroSchematic}b, can be considered to extract the constitutive responses through the computational homogenization \cite{KG:90}. 

\subsection{Macro- and microscale kinematics}
\label{sec:MacroMicroScaleKinematics}
\begin{figure}[!ht]
\centering
  \includegraphics[width=0.99\linewidth]{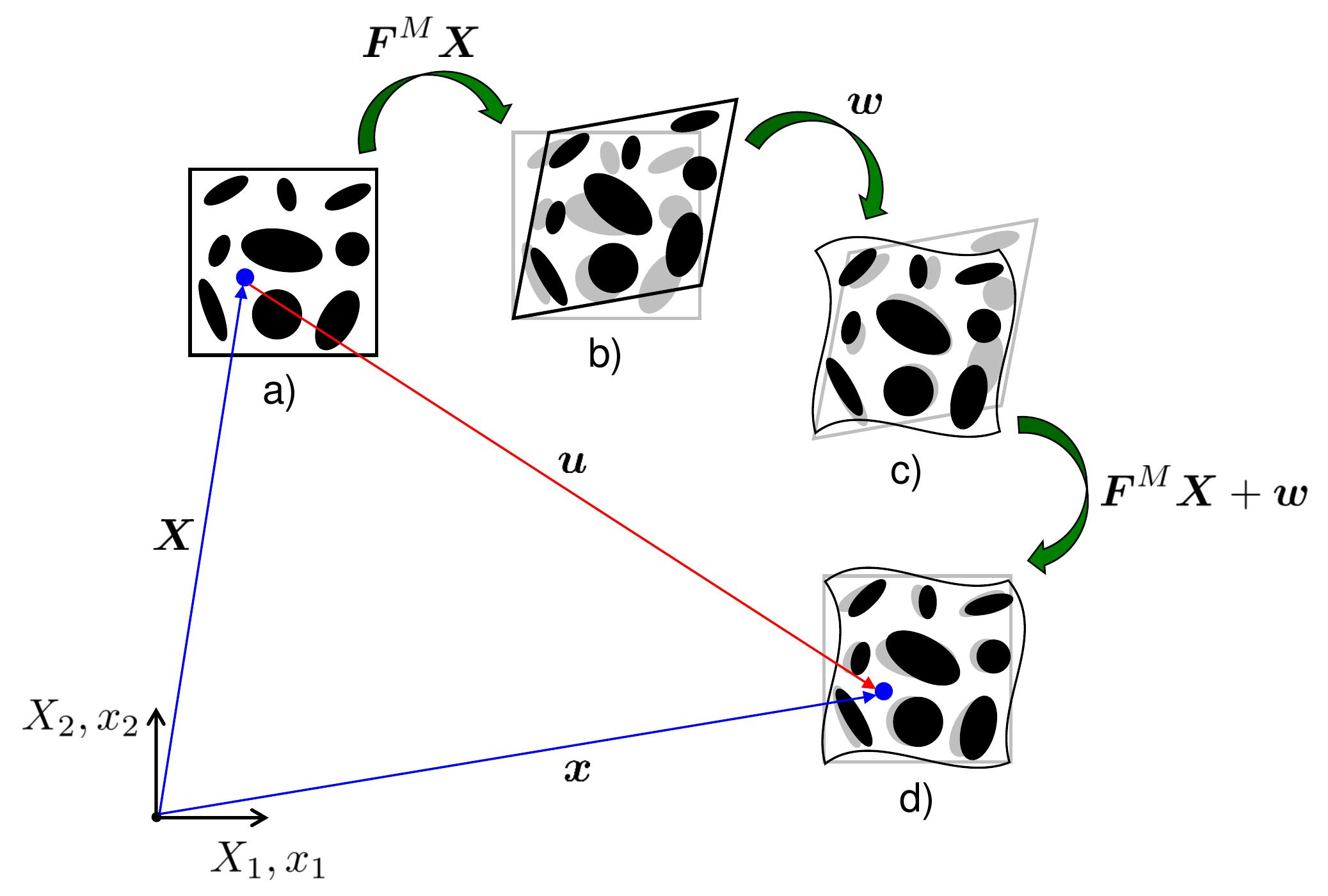}
\caption{Schematic representation of the deformation stages in the \ac{RUC}: a) reference configuration, b) the deformation caused by the macroscopic deformation gradient ($\boldsymbol{F}^{M} \boldsymbol{X}$), c) the microscopic fluctuation displacement ($\boldsymbol{w}$), and d) current configuration.}
\label{fig:02_MicroStructureDeformationSteps}
\end{figure}

Consider the \ac{RUC}, given in Fig. \ref{fig:01_MacroToMicroSchematic}c, that undergoes deformation. At both macro- and microscales, the deformation gradient, $\boldsymbol{F}$, the displacement of a material point, $\boldsymbol{u}$, and the displacement gradient, $\boldsymbol{G}$, are defined as follows:
\begin{equation}\label{eq:MacroMicroScaleKinematics_01}
\begin{split}
\boldsymbol{F} = \frac{\partial \boldsymbol{x}}{\partial \boldsymbol{X}}, \qquad \boldsymbol{u} = \boldsymbol{x} - \boldsymbol{X}, \qquad \boldsymbol{G} = \frac{\partial \boldsymbol{u}}{\partial \boldsymbol{X}} = \boldsymbol{F} - \boldsymbol{I},
\end{split}
\end{equation}
where $\boldsymbol{x}$ and $\boldsymbol{X}$ are the coordinates of material points in the current and reference configurations, respectively, and $\boldsymbol{I}$ is the second order identity tensor. Herein we follow the approach given in \cite{vanDijk:16}, where the deformation of the \ac{RUC} is decomposed into deformations caused by the macroscopic deformation and microscopic fluctuation displacement, as depicted in Fig. \ref{fig:02_MicroStructureDeformationSteps}. 

For a periodically patterned \ac{RVE}, the classical first-order homogenization theory can be used to relate the macroscopic deformation, $\boldsymbol{F}^{M}$, to the microscopic one, $\boldsymbol{F}^{\mu}$, \cite{KBB:01}:
\begin{equation}\label{eq:MacroMicroScaleKinematics_04}
\begin{split}
\boldsymbol{F}^{M} = \frac{1}{\Big| \Omega_{\mathrm{rve}} \Big|} \int_{\Omega_{\mathrm{rve}}} \boldsymbol{F}^{\mu} d\Omega,
\end{split}
\end{equation}
where $\Omega_{\mathrm{rve}}$ is the volume of the \ac{RVE}. A material point in the current configuration of the microscopic model, $\boldsymbol{x}$, can be related to the same material point in the reference configuration, $\boldsymbol{X}$, as follows \cite{KBB:01, vanDijk:16}:
\begin{equation}\label{eq:MacroMicroScaleKinematics_05}
\begin{split}
\boldsymbol{x} = \boldsymbol{F}^{M} \boldsymbol{X} + \boldsymbol{w},
\end{split}
\end{equation}
where $\boldsymbol{F}^{M} \boldsymbol{X}$ indicates the macroscopic deformation, Fig. \ref{fig:02_MicroStructureDeformationSteps}b. The macroscopic deformation gradient is constant over the \ac{RVE}. The variable $\boldsymbol{w}$ indicates the displacement caused by the microscopic fluctuation, Fig. \ref{fig:02_MicroStructureDeformationSteps}c.

\subsection{Periodic boundary conditions}
\label{sec:PeriodicBoundaryConditions}

\begin{figure}[t]
\centering
  \includegraphics[width=0.60\linewidth]{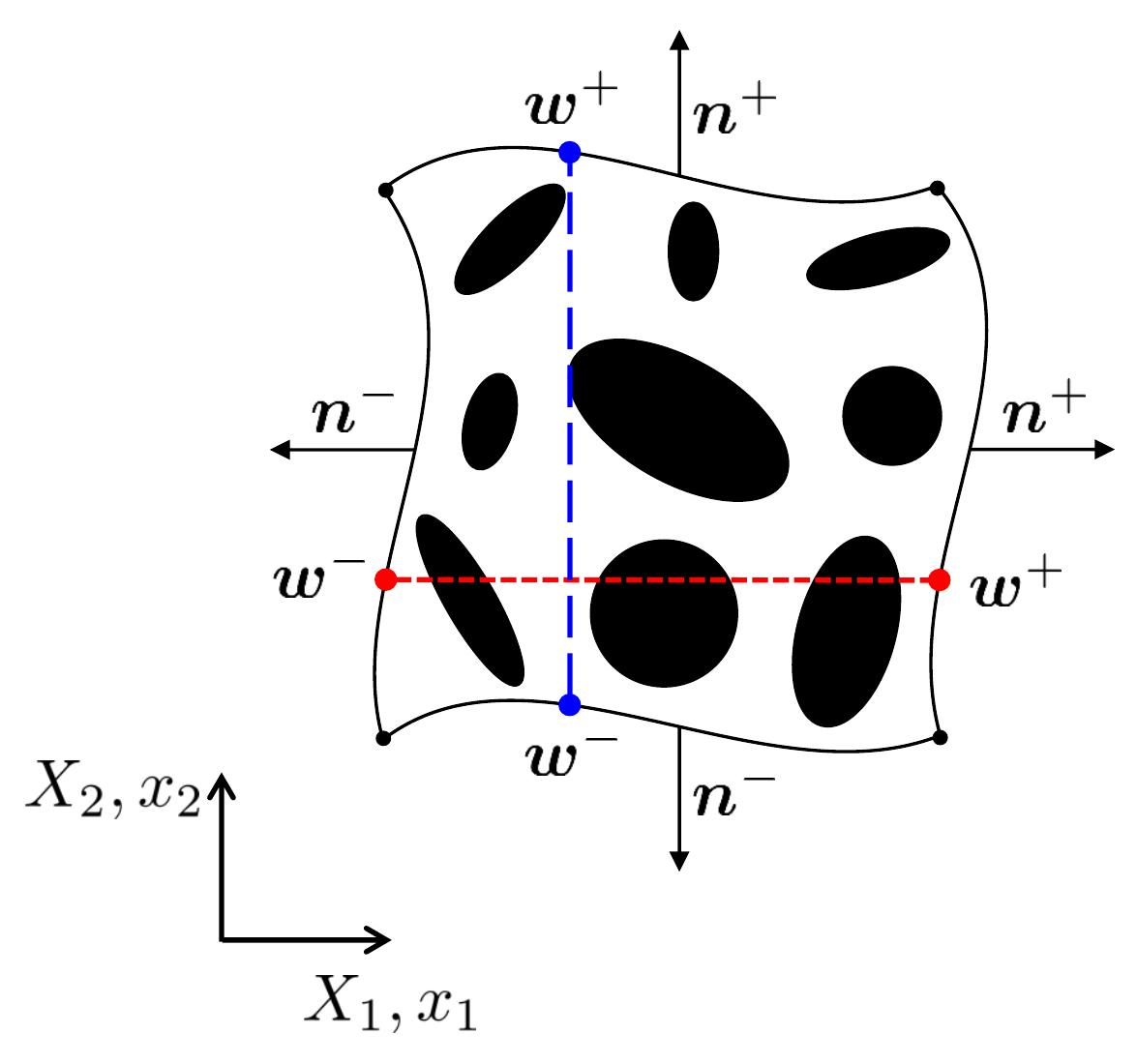}
\caption{Schematic representation of a 2D \ac{RVE} with \acp{PBC}. $\boldsymbol{w}^{-}$ and $\boldsymbol{w}^{+}$ correspond to microscopic fluctuation displacements on opposite boundaries, $\boldsymbol{n}^{-}$ and $\boldsymbol{n}^{+}$ are the outgoing normal vectors on opposite boundaries.}
\label{fig:03_PeriodicBCsSchematic}
\end{figure}
For the \ac{RVE}, the relationship between the macroscopic and microscopic deformation (Eq.~\eqref{eq:MacroMicroScaleKinematics_04}) can be simplified to the following boundary conditions \cite{KBB:01, vanDijk:16}:
\begin{equation}\label{eq:MacroMicroScaleKinematics_07}
\begin{split}
\frac{1}{\Big| \Omega_{\mathrm{rve}} \Big|} \int_{\Gamma_\mathrm{rve}} \boldsymbol{w} \cdot \boldsymbol{n} d\Gamma = \boldsymbol{0}, 
\end{split}
\end{equation}
where $\boldsymbol{n}$ is the outgoing normal vector over the boundaries of \ac{RVE}, as shown in Fig. \ref{fig:03_PeriodicBCsSchematic}. Eq.~\eqref{eq:MacroMicroScaleKinematics_07} indicates that the boundary conditions on the \ac{RVE} must be chosen such that the contribution of the microscale fluctuation displacement, $\boldsymbol{w}$, vanishes at the boundary. This requirement can be achieved in many ways \cite{GKM+:17}. \acp{PBC} could be an effective way to satisfy this requirement, where imposing the corresponding points on opposite boundaries (Fig. \ref{fig:03_PeriodicBCsSchematic}) yields:
\begin{equation}\label{eq:PeriodicBCsSchematic_01}
\begin{split}
\boldsymbol{w}^{+} = \boldsymbol{w}^{-}.
\end{split}
\end{equation}
For the numerical implementation of Eq.~\eqref{eq:PeriodicBCsSchematic_01}, it is assumed that the nodal distribution on the opposite boundaries of the \ac{RVE} are identical (Fig. \ref{fig:03_PeriodicBCsSchematic}). This assumption yields \enquote{a set of linearly independent boundary conditions} \cite{vanDijk:16}. For two spatial dimensions and $n$ nodes in the discretized 2D \ac{RVE}, the vector form of the microscopic fluctuation displacements can be constructed using Eqs.~\eqref{eq:MacroMicroScaleKinematics_01} and \eqref{eq:MacroMicroScaleKinematics_05} as follows:  
\begin{equation}\label{eq:PeriodicBCsSchematic_01_01}
\begin{split}
\underbrace{\boldsymbol{\widehat{w}}}_{2n \times 1} = \underbrace{\boldsymbol{\widehat{u}}}_{2n \times 1} - \underbrace{\boldsymbol{T}_{X}}_{2n \times 4} \underbrace{\boldsymbol{\widehat{G}}^{M}}_{4 \times 1},
\end{split}
\end{equation}
where $\boldsymbol{\widehat{u}}$ is the vector of microscale nodal displacement in the \ac{RVE} and $\boldsymbol{\widehat{G}}^{M}$ is the vector form of the macroscale displacement gradient (with size of $4 \times 1$ for 2D and $9 \times 1$ for 3D). The matrix $\boldsymbol{T}_{X}$ contains nodal coordinates of the discretized 2D \ac{RVE} (see Eq.~\eqref{eq:PeriodicBCs_Tx} in \ref{appx:HomogenMatrices}). Using Eqs.~\eqref{eq:PeriodicBCsSchematic_01_01}, the boundary conditions given in Eq.~\eqref{eq:PeriodicBCsSchematic_01} is expressed as \cite{vanDijk:16}:
\begin{equation}\label{eq:PeriodicBCsSchematic_01_02}
\begin{split}
\underbrace{\boldsymbol{T}_{p}}_{2m \times 2n} \Big( \underbrace{\boldsymbol{\widehat{u}}}_{2n \times 1} - \underbrace{\boldsymbol{T}_{X}}_{2n \times 4} \underbrace{\boldsymbol{\widehat{G}}^{M}}_{4 \times 1} \Big) = \mathbf{0},
\end{split}
\end{equation}
where $\boldsymbol{T}_{p}$ contains ($+1$) for positive nodes, ($-1$) for negative nodes on the boundaries, and zeros for nodes outside of the boundaries (Fig. \ref{fig:03_PeriodicBCsSchematic}). The number of rows (i.e., $2m$) in $\boldsymbol{T}_{p}$ denotes the number of required independent \ac{PBC} equations to be solved.

For homogenization where macroscopic strains are the degrees of freedom, \enquote{it is important to avoid the rotation part of the deformation gradient, as it would lead to a singular system of equations} \cite{vanDijk:16}. This can be achieved by modifying the polar decomposition of the deformation gradient tensor as follows:
\begin{equation}\label{eq:PeriodicBCsSchematic_01_02_1}
\begin{split}
\boldsymbol{F} = \boldsymbol{R} \boldsymbol{U}, \quad \text{with} \quad\boldsymbol{R} = \boldsymbol{I},
\end{split}
\end{equation}
where $\boldsymbol{R}$ is the rotation tensor and $\boldsymbol{U}$ is the symmetric right stretch tensor of the decomposed deformation gradient \cite{TN:04}, and $\boldsymbol{I}$ is the identity tensor. With the assumption of $\boldsymbol{R} = \boldsymbol{I}$, the macroscale displacement gradient can be expressed as:
\begin{equation}\label{eq:PeriodicBCsSchematic_01_03}
\begin{split}
\boldsymbol{\widehat{G}}^{M} = \boldsymbol{T}_{s} \boldsymbol{\widehat{G}}^{M}_{\mathrm{sym}},
\end{split}
\end{equation}
where $\boldsymbol{\widehat{G}}^{M}_{\mathrm{sym}}$ is the symmetric displacement gradient vector and $\boldsymbol{T}_{s}$ is an adjustment matrix (with zeros and ones) that recovers the full displacement gradient vector from the components of $\boldsymbol{\widehat{G}}^{M}_{\mathrm{sym}}$. Substituting Eq.~\eqref{eq:PeriodicBCsSchematic_01_03} into \eqref{eq:PeriodicBCsSchematic_01_02} leads to the following equations for \acp{PBC} that describe an implicit relationship between the microscale displacements, $\boldsymbol{\widehat{u}}$, and the macroscale Green-Lagrange strain, $\boldsymbol{\widehat{E}}^{M}$ \cite{vanDijk:16}:
\begin{equation}\label{eq:MacroMicroScaleStressStrainMeasures_08}
\begin{split}
\boldsymbol{T}_{p} \Big( \boldsymbol{\widehat{u}} - \boldsymbol{T}_{X} \boldsymbol{T}_{s} \boldsymbol{\widehat{G}}^{M}_{\mathrm{sym}} (\boldsymbol{\widehat{E}}^{M}) \Big) = \mathbf{0}.
\end{split}
\end{equation}
Additional details concerning the development of the \acp{PBC} are given in \ref{appx:HomogenMatrices} and can be found in \cite{vanDijk:16}.

\subsection{Microscale systems of equations}
\label{sec:MicroscaleSystemsofEquations}
Following \cite{vanDijk:16}, a Lagrange multiplier approach can be used to formulate the homogenization problem:
\begin{equation}\label{eq:LagrangeMultiplierApproach_01}
\begin{split}
\mathcal{L} (\boldsymbol{\widehat{u}}, \boldsymbol{\widehat{\lambda}}, \boldsymbol{\widehat{E}}^{M}) = \mathcal{W}^{M} (\boldsymbol{\widehat{u}}, \boldsymbol{\widehat{E}}^{M}) - \alpha \boldsymbol{\widehat{\lambda}}^{T} \boldsymbol{T}_{p} \Big( \boldsymbol{\widehat{u}} - \boldsymbol{T}_{X} \boldsymbol{T}_{s} \boldsymbol{\widehat{G}}^{M}_{\mathrm{sym}} (\boldsymbol{\widehat{E}}^{M}) \Big), 
\end{split}
\end{equation}
where $\boldsymbol{\widehat{\lambda}}$ is a vector of Lagrange multipliers, $\mathcal{W}^{M}$ is the macroscopic virtual work, and $\alpha$ is a positive numerical scaling factor that scales the constraint equations \enquote{for numerical conditioning of the global stiffness matrix} \cite{vanDijk:16}. In this paper, we set $\alpha = E$, where $E$ is the Young's modulus of the base material. This would scale the constraint equations to be the same order of magnitude of the elemental stiffness matrix. The equilibrium equations for the homogenization problem are \enquote{derived from the stationary condition} \cite{vanDijk:16} of Eq.~\eqref{eq:LagrangeMultiplierApproach_01} with respect to $\boldsymbol{\widehat{u}}$, $\boldsymbol{\widehat{\lambda}}$, $\boldsymbol{\widehat{E}}^{M}$. These derivations result in the following systems of equilibrium equations:
\begin{equation}\label{eq:LagrangeMultiplierApproach_02_00}
\begin{split}
\frac{\partial \mathcal{L}}{\partial \boldsymbol{\widehat{u}}} &= \mathbf{0}, \\ 
\frac{\partial \mathcal{L}}{\partial \boldsymbol{\widehat{\lambda}}} &= \mathbf{0}, \\ 
\frac{\partial \mathcal{L}}{\partial \boldsymbol{\widehat{E}}^{M}} &= \mathbf{0}.
\end{split}
\end{equation}
The derivative of $\mathcal{L}$ with respect to state variables (i.e., $\boldsymbol{\widehat{u}}$, $\boldsymbol{\widehat{\lambda}}$, $\boldsymbol{\widehat{E}}^{M}$) leads to the following system of residual equations \cite{vanDijk:16}:
\begin{equation}\label{eq:LagrangeMultiplierApproach_02}
\begin{split}
\boldsymbol{r} (\boldsymbol{\widehat{u}}, \boldsymbol{\widehat{\lambda}}, \boldsymbol{\widehat{E}}^{M})
=
\begin{pmatrix}
\displaystyle \frac{1}{\Big| \Omega_{\mathrm{rve}} \Big|} \boldsymbol{f}_{\mathrm{int}} (\boldsymbol{\widehat{u}}) - \alpha \boldsymbol{T}_{p}^{T} \boldsymbol{\widehat{\lambda}} \\ 
- \alpha \boldsymbol{T}_{p} \boldsymbol{\widehat{u}} + \alpha \boldsymbol{T}_{p} \boldsymbol{T}_{X} \boldsymbol{T}_{s} \boldsymbol{\widehat{G}}^{M}_{\mathrm{sym}} (\boldsymbol{\widehat{E}}^{M}) \\ 
\widehat{\boldsymbol{S}}^{M}_{\mathrm{int}} (\boldsymbol{\widehat{\lambda}}, \boldsymbol{\widehat{E}}^{M})
\end{pmatrix}
-
\begin{pmatrix}
\mathbf{0} \\ 
\mathbf{0} \\ 
\boldsymbol{\widehat{S}}^{M}
\end{pmatrix},
\end{split}
\end{equation}
where $\boldsymbol{f}_{\mathrm{int}}$ is the microscopic internal force vector, $\boldsymbol{\widehat{S}}^{M}$ is the macroscopic stress at a material point (i.e., applied macroscopic stress on the \ac{RVE}), and $\widehat{\boldsymbol{S}}^{M}_{\mathrm{int}}$ is the so-called internal macroscopic stress that corresponds to the macroscopic deformation of the \ac{RVE} given as follows:
\begin{equation}\label{eq:LagrangeMultiplierApproach_03}
\begin{split}
\widehat{\boldsymbol{S}}^{M}_{\mathrm{int}} (\boldsymbol{\widehat{\lambda}}, \boldsymbol{\widehat{E}}^{M}) = \alpha ( \boldsymbol{Z} (\boldsymbol{\widehat{E}}^{M}) )^{T} \boldsymbol{\widehat{\lambda}}, \quad \boldsymbol{Z} = \boldsymbol{T}_{p} \boldsymbol{T}_{X} \boldsymbol{T}_{s} {\boldsymbol{\bar M}}^{-1}, \quad \boldsymbol{\bar M} = \boldsymbol{\bar I} + \boldsymbol{\bar G}^{M}.
\end{split}
\end{equation}
The matrix $\boldsymbol{\bar G}^{M}$ collects the components of the symmetric displacement gradient and $\boldsymbol{\bar I}$ is a diagonal matrix \cite{vanDijk:16}. Detail on $\boldsymbol{\bar G}^{M}$ and $\boldsymbol{\bar I}$ is given in \ref{appx:HomogenMatrices}. 

\subsection{Homogenized tangent stiffness tensor}
\label{sec:HomogenizedTangentStiffnessTensor}
The macroscopic homogenized tangent stiffness tensor, $\mathbb{C}^{\mathrm{eff}}$, can be defined through the stress-strain relationship as follows \cite{vanDijk:16}:
\begin{equation}\label{eq:HomogenizedTangentStiffnessTensor_01}
\begin{split}
\delta \boldsymbol{\widehat{S}}^{M} = \mathbb{C}^{\mathrm{eff}} \delta \boldsymbol{\widehat{E}}^{M}.
\end{split}
\end{equation}
The tensor $\mathbb{C}^{\mathrm{eff}}$ can be computed from the converged solution of the microscopic equilibrium equations (i.e., Eq.~\eqref{eq:LagrangeMultiplierApproach_02}) as follows \cite{vanDijk:16}:
\begin{equation}\label{eq:HomogenizedTangentStiffnessTensor_02}
\begin{split}
\mathbb{C}^{\mathrm{eff}} (\boldsymbol{\widehat{u}},\boldsymbol{\widehat{\lambda}}, \boldsymbol{\widehat{E}}^{M})
&=
\mathbb{C}^{\mathrm{s}} (\boldsymbol{\widehat{\lambda}}, \boldsymbol{\widehat{E}}^{M}) \\
&-
\begin{bmatrix}
\mathbf{0} & \alpha \boldsymbol{Z}^{T} (\boldsymbol{\widehat{E}}^{M})
\end{bmatrix}
{\underbrace{
\begin{bmatrix}
\displaystyle \frac{1}{\Big| \Omega_{\mathrm{rve}} \Big|} \boldsymbol{K} (\boldsymbol{\widehat{u}}) & -\alpha \boldsymbol{T}_{p}^{T} \\
-\alpha \boldsymbol{T}_{p} & \mathbf{0}
\end{bmatrix}}_{\boldsymbol{\Upsilon}}}^{-1}
\begin{bmatrix}
\mathbf{0} \\ 
\alpha \boldsymbol{Z} (\boldsymbol{\widehat{E}}^{M})
\end{bmatrix},
\end{split}
\end{equation}
where $\boldsymbol{K} = \partial \boldsymbol{f}_{\mathrm{int}}^{\mu} (\boldsymbol{\widehat{u}}) / \partial \boldsymbol{\widehat{u}}$ is the microscale tangent stiffness matrix and $\mathbb{C}^{\mathrm{s}}$ is a type of geometrical stiffness tensor given as:
\begin{equation}\label{eq:LagrangeMultiplierApproach_07}
\begin{split}
\mathbb{C}^{\mathrm{s}} = - {\boldsymbol{\bar M}}^{-1} \boldsymbol{\bar S}^{M} {\boldsymbol{\bar M}}^{-1},
\end{split}
\end{equation}
with $\boldsymbol{\bar S}^{M}$ derived from the components of $\widehat{\boldsymbol{S}}^{M}_{\mathrm{int}}$ (defined in Eq.~\eqref{eq:LagrangeMultiplierApproach_03}), see \ref{appx:HomogenMatrices}.

To facilitate the complex derivations of $\mathbb{C}^{\mathrm{eff}}$ used for sensitivity analysis purposes, we simplified Eq.~\eqref{eq:HomogenizedTangentStiffnessTensor_02} by modifying the inverse of $\boldsymbol{\Upsilon}$ matrix in the following setting. Consider a $2 \times 2 $ block of matrices given as follows:
\begin{equation}
\label{eq:LemmaBlockMatrices_01}
\begin{split}
\boldsymbol{R}
=
\begin{bmatrix}
\boldsymbol{A}_{k \times m} & \boldsymbol{B}_{k \times n} \\
\boldsymbol{C}_{l \times m} & \boldsymbol{D}_{l \times n}
\end{bmatrix}_{(k+l) \times (m+n)},
\end{split}
\end{equation}
where $\boldsymbol{R}$ is a square matrix with $k + l = m + n$. For the case when $\boldsymbol{A}$ is a non-singular square matrix (i.e., $\boldsymbol{A}^{-1} \neq 0$ and $k = m$) and $\boldsymbol{D}$ is a square matrix (i.e., $l = n$), the square matrix $\boldsymbol{R}$ is invertible if and only if the Schur complement (i.e., $\boldsymbol{D} - \boldsymbol{C}\boldsymbol{A}^{-1} \boldsymbol{B}$) of $\boldsymbol{A}$ is invertible \cite{LS:02}. For this case, the inverse of $\boldsymbol{R}$ is given as follows \cite{LS:02}:
\begin{equation}
\label{eq:LemmaBlockMatrices_02}
\begin{split}
\boldsymbol{R}^{-1}
=
\begin{bmatrix}
\boldsymbol{A}^{-1} + \boldsymbol{A}^{-1} \boldsymbol{B} (\boldsymbol{D} - \boldsymbol{C}\boldsymbol{A}^{-1} \boldsymbol{B})^{-1} \boldsymbol{C} \boldsymbol{A}^{-1} &  -\boldsymbol{A}^{-1} \boldsymbol{B} (\boldsymbol{D} - \boldsymbol{C}\boldsymbol{A}^{-1} \boldsymbol{B})^{-1} \\
-(\boldsymbol{D} - \boldsymbol{C}\boldsymbol{A}^{-1} \boldsymbol{B})^{-1} \boldsymbol{C} \boldsymbol{A}^{-1} & (\boldsymbol{D} - \boldsymbol{C}\boldsymbol{A}^{-1} \boldsymbol{B})^{-1}
\end{bmatrix}.
\end{split}
\end{equation}
The $\boldsymbol{\Upsilon}$ matrix given in Eq.~\eqref{eq:HomogenizedTangentStiffnessTensor_02} contains a $2 \times 2$ block of matrices where its internal matrices have the aforementioned properties with:
\begin{equation}
\label{eq:LemmaBlockMatrices_03}
\begin{split}
\boldsymbol{A} = \frac{1}{\Big| \Omega_{\mathrm{rve}} \Big|} \boldsymbol{K}, \quad \boldsymbol{B} = \boldsymbol{C}^{T} = -\alpha \boldsymbol{T}_{p}^{T}, \quad \boldsymbol{D} = \mathbf{0},
\end{split}
\end{equation}
where the stiffness matrix, $\boldsymbol{K}$, is a non-singular square matrix, $\boldsymbol{D}$ is a square matrix, and the Schur complement of $\boldsymbol{A}$ is invertible; therefore, one can conclude that $\boldsymbol{\Upsilon}^{-1} = \boldsymbol{R}^{-1}$. The use of Eqs.~\eqref{eq:LemmaBlockMatrices_02} and \eqref{eq:LemmaBlockMatrices_03} in \eqref{eq:HomogenizedTangentStiffnessTensor_02} yields the simplified form of the homogenized stiffness tensor as follows:
\begin{equation}
\label{eq:LemmaBlockMatrices_04}
\begin{split}
\mathbb{C}^{\mathrm{eff}}
=
\mathbb{C}^{\mathrm{s}}
-
\boldsymbol{Z}^{T}  \boldsymbol{\Psi}^{-1} \boldsymbol{Z}, \qquad \text{with} \qquad \boldsymbol{\Psi} = - \Big| \Omega_{\mathrm{rve}} \Big| \boldsymbol{T}_{p} \boldsymbol{K} ^{-1} \boldsymbol{T}_{p}^{T}.
\end{split}
\end{equation}
%


%
\section{Sensitivity Analysis}
\label{sec:SensitivityAnalysis}
\subsection{Adjoint Sensitivity}
\label{sec:AdjointSensitivity}
Adjoint-based sensitivities are derived to compute the sensitivity of a criterion (the objective or a constraint) with respect to a design-dependent fictitious elemental density, $\rho^{e}$, as follows:
\begin{equation}\label{eq:SensitivityAnalysis_03}
\begin{split}
\frac{d \mathcal{Z}}{d \rho^{e}} = \frac{\partial \mathcal{Z}}{\partial \rho^{e}} + \boldsymbol{\tilde \chi}^{T} \frac{\partial \boldsymbol{r}}{\partial \rho^{e}},
\end{split}
\end{equation}
where $\mathcal{Z}$ is defined as the objective function or a constraint, and $\boldsymbol{\tilde \chi}$ is the vector of adjoint solutions, computed as follows:
\begin{equation}\label{eq:SensitivityAnalysis_04}
\begin{split}
\left (\frac{\partial \boldsymbol{r}}{\partial \boldsymbol{s}}  \right )^{T} \boldsymbol{\tilde \chi} = - \left ( \frac{\partial \mathcal{Z}}{\partial \boldsymbol{s}} \right )^{T},
\end{split}
\end{equation}
where $\boldsymbol{s} = [\boldsymbol{\widehat{u}}, \boldsymbol{\widehat{\lambda}}, \boldsymbol{\widehat{E}}^{M}]^{T}$ is the vector of state variables satisfying the microscale equilibrium. The derivatives of the residuals with respect to state variables are computed using Eq.~\eqref{eq:LagrangeMultiplierApproach_02}. For a generalized optimization problem, a criterion could be an explicit function of the homogenized tangent stiffness tensor, $\mathbb{C}^{\mathrm{eff}}$. Hence, the detail on the derivation of $\mathbb{C}^{\mathrm{eff}}$ with respect to the design and state variables are given below.

\subsection{Derivative of $\mathbb{C}^{\mathrm{eff}}$ with respect to the elemental density}
\label{sec:dCeff_drho}
The partial derivative of $\mathbb{C}^{\mathrm{eff}}$ with respect to $\rho^{e}$ are derived from Eq.~\eqref{eq:LemmaBlockMatrices_04} as follows:
\begin{equation}
\label{eq:dCeff_drho_03}
\begin{split}
\frac{\partial \mathbb{C}^{\mathrm{eff}}}{\partial \rho^{e}}
=
\boldsymbol{Z}^{T} \boldsymbol{\Psi}^{-1} \frac{\partial \boldsymbol{\Psi}}{\partial \rho^{e}} \boldsymbol{\Psi}^{-1} \boldsymbol{Z}, \quad \text{with} \quad \frac{\partial \boldsymbol{\Psi}}{\partial \rho^{e}} = \Big| \Omega_{\mathrm{rve}} \Big| \boldsymbol{T}_{p} \boldsymbol{K}^{-1} \frac{\partial \boldsymbol{K}}{\partial \rho^{e}} \boldsymbol{K}^{-1} \boldsymbol{T}_{p}^{T}.
\end{split}
\end{equation}
We note that $\mathbb{C}^{\mathrm{s}}$ and $\boldsymbol{Z}$ (in Eq.~\eqref{eq:LemmaBlockMatrices_04}) are independent of $\rho^{e}$. The term $\partial \boldsymbol{K}/\partial \rho^{e}$ can be computed from the material interpolation scheme given in Section \ref{sec:MaterialInterpolation}.  

\subsection{Derivative of $\mathbb{C}^{\mathrm{eff}}$ with respect to microscale displacements}
\label{sec:dCeff_dup}
From Eq.~\eqref{eq:LemmaBlockMatrices_04}, the partial derivative of $\mathbb{C}^{\mathrm{eff}}$ with respect to a displacement \ac{DOF}, $\widehat{u}_{p}$, is given as follows:
\begin{equation}
\label{eq:dCeff_dup_01}
\begin{split}
\frac{\partial \mathbb{C}^{\mathrm{eff}}}{\partial \widehat{u}_{p}}
=
\boldsymbol{Z}^{T} \boldsymbol{\Psi}^{-1} \frac{\partial \boldsymbol{\Psi}}{\partial \widehat{u}_{p}} \boldsymbol{\Psi}^{-1} \boldsymbol{Z}, \quad \text{with} \quad \frac{\partial \boldsymbol{\Psi}}{\partial \widehat{u}_{p}} = \Big| \Omega_{\mathrm{rve}} \Big| \boldsymbol{T}_{p} \boldsymbol{K}^{-1} \frac{\partial \boldsymbol{K}}{\partial \widehat{u}_{p}} \boldsymbol{K}^{-1} \boldsymbol{T}_{p}^{T}.
\end{split}
\end{equation}
The detail on the first and second derivatives of the microscale internal strain energy is given in \ref{appx:DerivativeofStiffnessMatrix}. The derivations consider both material and geometric nonlinearities. 

\subsection{Derivative of $\mathbb{C}^{\mathrm{eff}}$ with respect to Lagrange multipliers}
\label{sec:dCeff_dlambdar}
The derivative of $\mathbb{C}^{\mathrm{eff}}$, i.e., Eq.~\eqref{eq:LemmaBlockMatrices_04}, with respect to a Lagrange multiplier, $\widehat{\lambda}_{r}$, is given as:
\begin{equation}
\label{eq:dCeff_dlambdar_01}
\begin{split}
\frac{\partial \mathbb{C}^{\mathrm{eff}}}{\partial \widehat{\lambda}_{r}}
&=
- {\boldsymbol{\bar M}}^{-1} \frac{\partial \boldsymbol{\bar S}^{M}}{\partial \widehat{\boldsymbol{S}}^{M}_{\mathrm{int}}} \frac{\partial \widehat{\boldsymbol{S}}^{M}_{\mathrm{int}}}{\partial \lambda_{r}}
 {\boldsymbol{\bar M}}^{-1}.
\end{split}
\end{equation}
The term $\partial \widehat{\boldsymbol{S}}^{M}_{\mathrm{int}}/ \partial \lambda_{r}$ can be computed from Eq.~\eqref{eq:LagrangeMultiplierApproach_03} and the derivatives of $\boldsymbol{\bar S}^{M}$ with respect to internal stress components can be computed from Eq.~\eqref{eq:LagrangeMultiplierApproach_08}.

\subsection{Derivative of $\mathbb{C}^{\mathrm{eff}}$ with respect to macroscale strains}
\label{sec:dCeff_dEr}
The derivative of $\mathbb{C}^{\mathrm{eff}}$ with respect to a macroscale strain \ac{DOF}, $\widehat{E}^{M}_{r}$, can be computed from Eq.~\eqref{eq:LemmaBlockMatrices_04} as follows:
\begin{equation}
\label{eq:dCeff_dEr_01}
\begin{split}
\frac{\partial \mathbb{C}^{\mathrm{eff}}}{\partial \widehat{E}^{M}_{r}}
&=
\frac{\partial \mathbb{C}^{\mathrm{s}}}{\partial \widehat{E}^{M}_{r}}
-
\Big( \frac{\partial \boldsymbol{Z}^{T}}{\partial \widehat{E}^{M}_{r}} \boldsymbol{\Psi}^{-1} \boldsymbol{Z} + \boldsymbol{Z}^{T} \boldsymbol{\Psi}^{-1} \frac{\partial \boldsymbol{Z}}{\partial \widehat{E}^{M}_{r}} \Big),
\end{split}
\end{equation}
where the derivative of $\mathbb{C}^{\mathrm{s}}$ and $\boldsymbol{Z}$ with respect to macroscale strain \acp{DOF} are computed from Eqs.~\eqref{eq:LagrangeMultiplierApproach_03} and \eqref{eq:LagrangeMultiplierApproach_07} as follows:
\begin{equation}
\label{eq:dCeff_dEr_02}
\begin{split}
\frac{\partial \mathbb{C}^{\mathrm{s}}}{\partial \widehat{E}^{M}_{r}}
&=
- \Bigg( \frac{\partial \boldsymbol{\bar M}^{-1}}{\partial \widehat{E}^{M}_{r}} \boldsymbol{\bar S}^{M} \boldsymbol{\bar M}^{-1} + \boldsymbol{\bar M}^{-1} \frac{\partial \boldsymbol{\bar S}^{M}}{\partial \widehat{E}^{M}_{r}}
 \boldsymbol{\bar M}^{-1} + \boldsymbol{\bar M}^{-1} \boldsymbol{\bar S}^{M} \frac{\partial \boldsymbol{\bar M}^{-1}}{\partial \widehat{E}^{M}_{r}} \Bigg), \\
\frac{\partial \boldsymbol{Z}}{\partial \widehat{E}^{M}_{r}}
&=
- \boldsymbol{Z} \frac{\partial \boldsymbol{\bar G}^{M}}{\partial \boldsymbol{\widehat{G}}^{M}_{\mathrm{sym}}} \boldsymbol{\bar M}^{-1} \boldsymbol{\bar M}^{-1}, 
\end{split}
\end{equation}
with
\begin{equation}
\label{eq:dCeff_dEr_03}
\begin{split}
\frac{\partial \boldsymbol{\bar M}^{-1}}{\partial \widehat{E}^{M}_{r}}
=
- \boldsymbol{\bar M}^{-1} \frac{\partial \boldsymbol{\bar G}^{M}}{\partial \boldsymbol{\widehat{G}}^{M}_{\mathrm{sym}}} \boldsymbol{\bar M}^{-1} \boldsymbol{\bar M}^{-1}.
\end{split}
\end{equation}
The derivatives of $\boldsymbol{\bar G}^{M}$ with respect to the components of $\boldsymbol{\widehat{G}}^{M}_{\mathrm{sym}}$ can be computed from Eq.~\eqref{eq:MacroMicroScaleStressStrainMeasures_06}.


%
\section{Topology Optimization}
\label{sec:TopologyOptimization}
The optimization problem presented here is included as an example of how to consider the components of the homogenized tangent stiffness tensor, $\mathbb{C}^{\mathrm{eff}}$, in the formulation of the objective function and constraints. Many combinations of individual components of the homogenized tangent stiffness tensor can be considered (for instance, to maximize the material bulk modulus \cite{Sigmund:00, AGN+:10} or Poisson's ratio \cite{ALS:14}). In this paper, the objective is to minimize the difference between the computed and target tangent stiffness tensors. The generalized formulation of the objective and constraints, material interpolation, and regularization are discussed below. 

\subsection{Objective and Constraints}
\label{sec:ObjectiveAndConstraints}
The optimization problem for minimizing the difference between the homogenized tangent stiffness tensor, $\mathbb{C}^{\mathrm{eff}}$, and the target tangent stiffness tensor, $\mathbb{C}^{\mathrm{target}}$, is formulated as follows:
\mathleft
\begin{equation}\label{eq:OptimizationProblem_01}
\begin{split}
\underset{\boldsymbol{\phi}}{\text{min}} & \qquad z = \displaystyle \sum_{i,j,k,l = 1}^{D} \Big( \mathbb{C}^{\mathrm{eff}}_{ijkl} ( \boldsymbol{s} (\boldsymbol{\phi}), \boldsymbol{\phi} )  - \mathbb{C}^{\mathrm{target}}_{ijkl} \Big)^2 & \\
\text{subject to} &
\begin{cases}
\boldsymbol{r} (\boldsymbol{s} (\boldsymbol{\phi}), \boldsymbol{\phi}) = \boldsymbol{0} \\
\displaystyle \sum_{e \in \Omega_{\mathrm{rve}}} \rho^{e} (\boldsymbol{\phi}) v^{e} \leq V_{max} \\ 
0 \leq \phi_{k} \leq 1 \qquad \forall k = 1, ..., N_{\phi}
\end{cases}
&
\end{split}
,
\end{equation}
\mathcenter

\noindent
where $z$ is the objective function, $D$ is the spatial dimension, $\boldsymbol{\phi}$ is the vector of independent design variables, $\rho^{e}$ is the fictitious elemental density, $v^{e}$ is the elemental volume, and $V_{max}$ is the prescribed volume fraction of the solid material. The \ac{NAND} approach is used to solve the microscale finite element problem where only the design variables, $\boldsymbol{\phi}$, are treated as the independent optimization variables \cite{AW:05}. The independent design variables are bounded with lower and upper bounds with $N_{\phi}$ that denotes the number of design variables.

\subsection{Material Interpolation}
\label{sec:MaterialInterpolation}
The \ac{SIMP} approach \cite{Bendsoe:89} is used for the interpolation of material properties with an artificial power law that penalizes intermediate density values. Using this approach, the stiffness of the microstructure is related to topology through the design dependent Young's modulus as follows:
\begin{equation}\label{eq:MaterialInterpolation_01}
\begin{split}
E^{e} (\rho^{e}) = \Big( \rho_{min}^{e} + (\rho^{e})^{\bar{p}} \big(1 - \rho_{min}^{e}\big) \Big) E_{0}^{e},
\end{split}
\end{equation}
where $E_{0}^{e}$ is the Young's modulus of pure solid material, $\bar{p} \geq 1$ is the \ac{SIMP} exponent penalty term, and $\rho_{min}^{e}$ is a small positive number to maintain positive definiteness of the global stiffness matrix. We set $\rho^{e} = 1$ for the solid region and $\rho^{e} = 0$ for the void region.

\subsection{Regularization}
\label{sec:Regularization}
To circumvent instabilities caused by checkerboard patterns \cite{SP:98}, and convergence to binary solutions, numerical regularization approaches such as filtering \cite{BT:01} and projection \cite{GPB:04, Sigmund:07, WLS:11} are commonly used in topology optimization. In this paper, we adopt the linear density filtering \cite{BT:01} and the threshold projection \cite{WLS:11} approaches. Our computational experiment showed that the threshold projection approach results in a more stable convergence behavior and yields discrete designs \cite{WLS:11, WSJ:14}. To eliminate the checkerboard patterns, the independent nodal design variables field, $\boldsymbol{\phi}$, is regularized using a linear filtering scheme as follows:
\begin{equation}\label{eq:Regularization_01}
\begin{split}
\mu^{e} (\boldsymbol{\phi}) = \frac{\displaystyle \sum_{i \in N^{e}} \hat{w}_{i} \phi_{i}}{\displaystyle \sum_{i \in N^{e}} \hat{w}_{i}}, \quad \text{with} \quad \hat{w}_{i} = \frac{r_{min} - \parallel x_{i} - x^{e} \parallel}{r_{min}},
\end{split}
\end{equation}
where $N^{e}$ contains all design variables located within a radius $r_{min}$, $x_{i}$ and $x^{e}$ are the position of node $i$ and the central position of element $e$, respectively. The elemental density is then related to the regularized design variables through the threshold projection where different design realizations for the manufacturing process can be considered by choosing different thresholds, $\eta$:
\begin{equation}\label{eq:Regularization_02}
\begin{split}
\rho^{e} (\boldsymbol{\phi}) = \frac{\tanh(\beta \eta) + \tanh(\beta( \mu^{e} (\boldsymbol{\phi}) - \eta) )}{\tanh(\beta \eta) + \tanh(\beta( 1 - \eta) )},
\end{split}
\end{equation}
where $\beta > 0 $ dictates the curvature of the regularization which approaches the Heaviside function as $\beta \to \infty$ \cite{WLS:11}.

Accounting for the filtering and projection (i.e., Eqs.~\eqref{eq:Regularization_01} and \eqref{eq:Regularization_02}), the derivative of the objective function and the residual (given in Eq.~\eqref{eq:SensitivityAnalysis_03}) with respect to an independent design variable ($\phi_{k}$) is given through the chain rule as follows:
\begin{equation}\label{eq:SensitivityAnalysisChain_01}
\begin{split}
\frac{\partial \mathcal{Z}}{\partial \phi_{k}} = \sum_{e \in N^{h}} \frac{\partial \mathcal{Z}}{\partial \rho^{e}} \frac{\partial \rho^{e}}{\partial \mu^{e}} \frac{\partial \mu^{e}}{\partial \phi_{k}}, \\
\frac{\partial \boldsymbol{r}}{\partial \phi_{k}}= \sum_{e \in N^{h}} \frac{\partial \boldsymbol{r}}{\partial \rho^{e}} \frac{\partial \rho^{e}}{\partial \mu^{e}} \frac{\partial \mu^{e}}{\partial \phi_{k}},
\end{split}
\end{equation}
where $N^{h}$ defines set containing all elements located within the distance $r_{min}$ of design variables $k$. The second and third part of the derivation in Eq.~\eqref{eq:SensitivityAnalysisChain_01} (i.e., $\frac{\partial \rho^{e}}{\partial \mu^{e}}$ and $\frac{\partial \mu^{e}}{\partial \phi_{k}}$) can be computed from Eqs.~\eqref{eq:Regularization_01} and \eqref{eq:Regularization_02}.
%

\subsection{Algorithmic Summary}
\label{sec:AlgorithmicSummary}
For the design problems presented in this paper, the \ac{MMA} optimizer \cite{Svanberg:87} is used, which has proven an efficient and reliable gradient-based optimization engine for solving a wide range of nonlinear structural optimization problems \cite{SM:13}. The most relevant \ac{MMA} parameters are summarized in Table \ref{tab:MMAparameters}. At each design iteration, the microscale nonlinear finite element problem is solved using the arc-length method \cite{DCR+:12}. We note that the arc-length method is very efficient in solving nonlinear systems of equations when the problem under consideration exhibits one or more critical instabilities due to an increase of the external forces or a problem undergoes large deformation. The finite element problem is considered converged if the relative change of the residuals (Eq.~\eqref{eq:LagrangeMultiplierApproach_02}) is less than $10^{-10}$. The solution for the adjoint variables given in Eq.~\eqref{eq:SensitivityAnalysis_04} is obtained using the direct linear solver with lower-upper (LU) factorization \cite{GMS:92}. We note that for a large scale system of equations, the direct linear solver might be inefficient and costly, and iterative methods [e.g., \ac{GMRES} \cite{SS:86}] can instead be used. To avoid convergence to undesirable local minima with an oscillatory design history, continuations on the $\beta$ parameter (Eq.~\eqref{eq:Regularization_02}) are considered \cite{WLS:11}. The optimization problem is considered converged if all constraints are satisfied and the relative change of the objective function is less than $10^{-6}$.
\begin{table}[!t]
\centering
		\caption{\ac{MMA} parameters used in topology optimization problems.}
		\label{tab:MMAparameters}      
		\begin{tabular}{@{}llll@{}}
			\hline
			Description & Symbol & Value \\
			\hline
			initial asymptote parameter & $\sigma_{\mathrm{MMA}}$ & $0.017$ \\
			lower asymptote adaptivity & $\alpha_{-}$& $0.55$ \\
			upper asymptote adaptivity & $\alpha_{+}$& $1.05$ \\
			constraint penalty & $c_{i}$ & $1000$ \\
			\hline
		\end{tabular}
\end{table}
%

%
\section{Numerical examples}
\label{sec:NumericalExamples}
In this section, we implement the topology optimization approach in 2D, assuming plane stress state, for uniaxially applied strain. For all numerical examples, the periodic microstructured \ac{RVE} given in Fig. \ref{fig:01_MacroToMicroSchematic}c is considered, where $l_{1} = l_{2} = 1$. The domain is discretized with $100 \times 100$ bilinear quadrilateral elements. We note that our mesh refinement studies show negligible discretization errors. Symmetry on the design is enforced in the axial directions and along the diagonals (Fig. \ref{fig:01_MacroToMicroSchematic}c). To eliminate the rigid-body translation, an arbitrary node in the solid region of the design domain (shaded area in Fig. \ref{fig:01_MacroToMicroSchematic}c) of the microscale problem is constrained in all directions and reflected based on the applied axes of symmetry. The base material parameters are set to $E = 1$ and $\nu = 0.3$.

\subsection{Topology optimized designs}
\label{sec:TODesignsBiaxialLoading}
\begin{figure}[t]
\centering
  \includegraphics[width=0.60\linewidth]{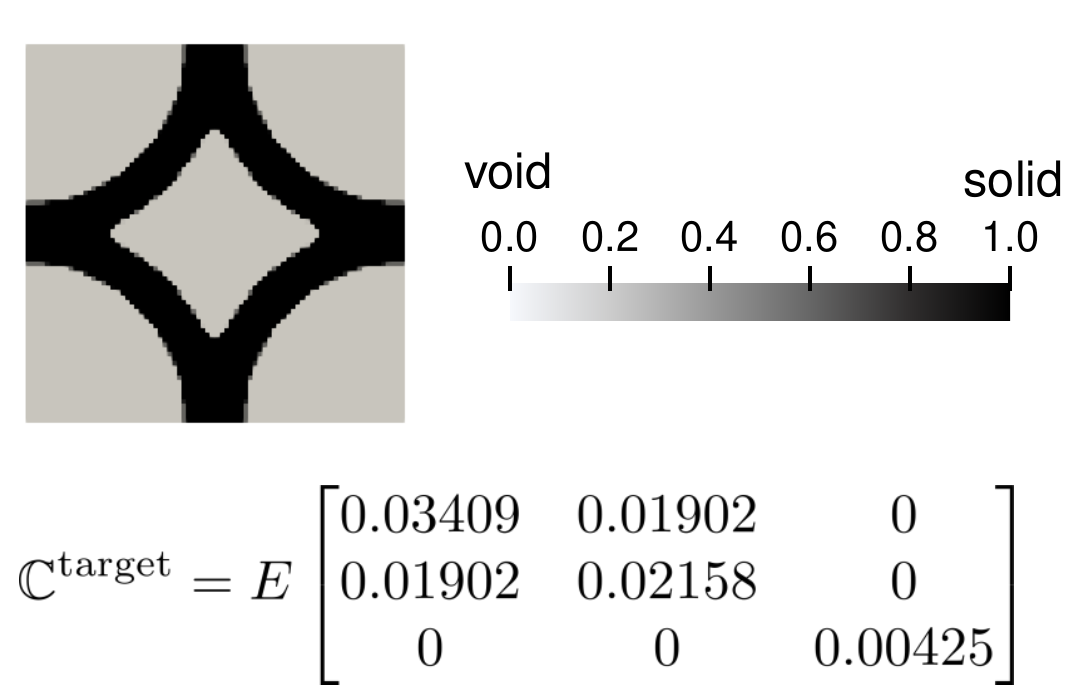}
\caption{The target stiffness tensor ($\mathbb{C}^{\mathrm{target}}$) used in topology optimization problems and the corresponding geometry, which was used to seed the target stiffness tensor. $E$ is the Young's modulus of the base material.}
\label{fig:04_Target_02_05_3_rmin_2}
\end{figure}

\begin{figure}[!ht]
\centering
  \includegraphics[width=0.90\linewidth]{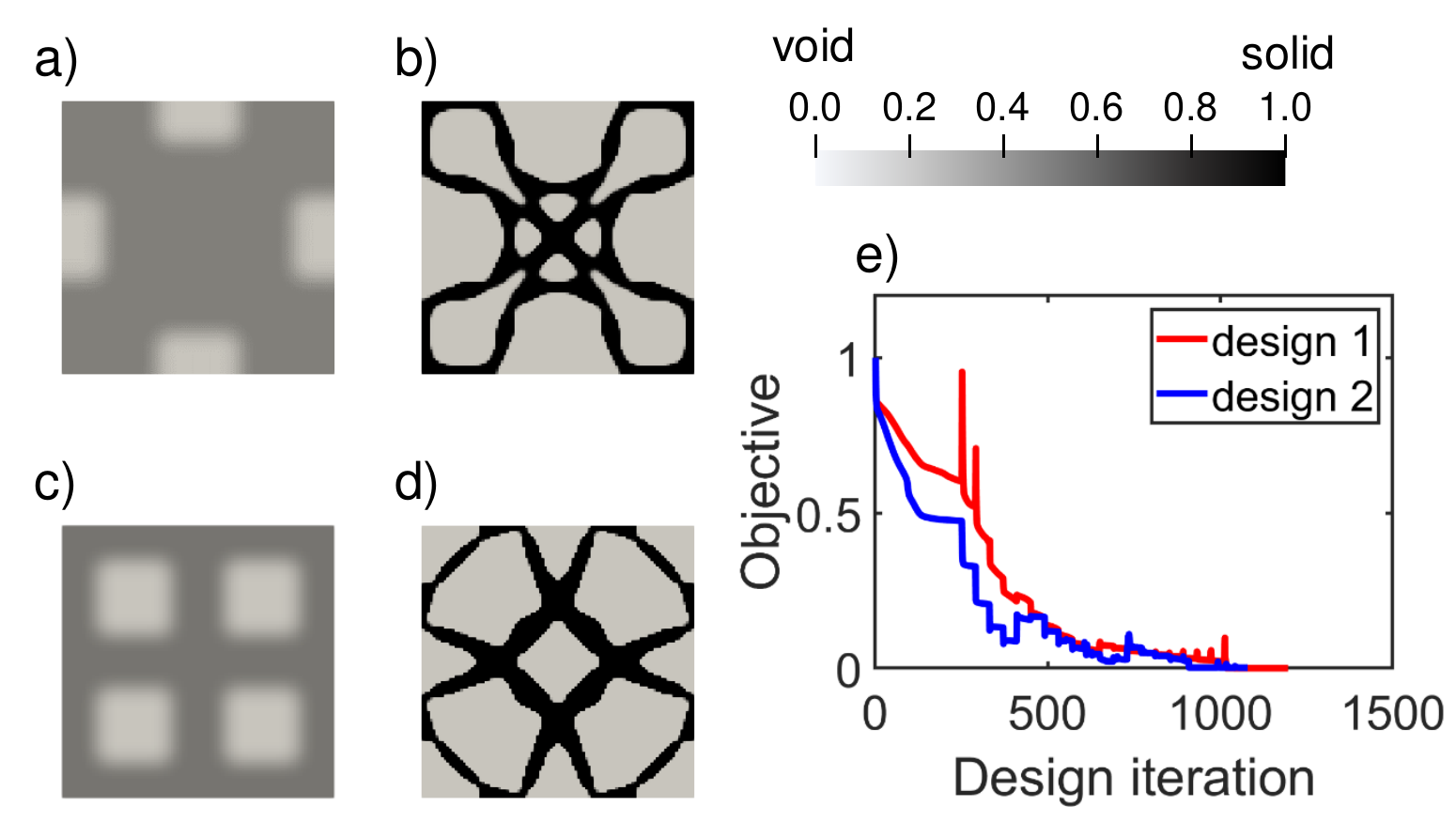}
\caption{a) Initial guess for design 1. b) Optimized design for design 1. c) Initial guess for design 2. d) Optimized design for design 2. e) The objective history for design 1 and 2.}
\label{fig:05_Design_02_05_3_rmin_2}
\end{figure}

In the section, we present an example using this optimization framework where we start with two different initial guesses, and seek to match the target stiffness tensor, i.e., $\mathbb{C}^{\mathrm{target}}$ at 20\% strain. We note that an arbitrary geometry given in Fig. \ref{fig:04_Target_02_05_3_rmin_2} is used to seed the target stiffness tensor. Table \ref{tab:InfluenceofInitialGuess} summarizes the parameters defining the optimization problem. The minimum desirable features are set equal to the minimum feature of the geometry given in Fig. \ref{fig:04_Target_02_05_3_rmin_2} through the filter radius as given in Table \ref{tab:InfluenceofInitialGuess}. The resulting optimized microstructures are shown in Fig. \ref{fig:05_Design_02_05_3_rmin_2}b and Fig. \ref{fig:05_Design_02_05_3_rmin_2}d. The results reveal the influence of the initial guess on the response of the topology optimization designs. For all designs, the target stiffness tensor is achieved, satisfying the prescribed volume constraint. While multiple local minima are obtained, they are equivalently good. The evolution of the objective function is shown in Fig. \ref{fig:05_Design_02_05_3_rmin_2}e. The reported maximum discrepancy between the optimized and the target responses is less than 0.9\%. The oscillations in the objective evolution are caused by the continuation in the projection parameter, $\beta$, where an update in $\beta$ results in the change in the elemental density and eventually the physical response and the objective function. Although the topology optimized designs qualitatively match the target response, the results reveal that the optimized design is not unique and for the same physical response multiple optimized designs can be obtained.
\begin{table}[t]
\centering
		\caption{Topology optimization and finite element modeling parameters for the design problems.}
		\label{tab:InfluenceofInitialGuess}      
		\begin{tabular}{@{}llll@{}}
			\hline
			Description & Symbol & Value \\
			\hline
			applied macroscale strain & ${\widehat{E}}^{M}_{yy}$ & 20\% \\
			\ac{RVE} thickness & -- & $0.3$ \\
			target volume fraction & $V_{max}$ & $0.305$ \\
			filter radius & $r_{min}$ & $0.0875$ \\
			\ac{SIMP} exponent & $\bar{p}$ & $3$ \\
			small positive number & $\rho^{e}_{min}$ & $10^{-4}$ \\
			threshold of the projection & $\eta$ & 0.50 \\
			curvature of regularization & $\beta$ & $2 \leq \beta \leq 100$ \\
			\hline
		\end{tabular}
\end{table}
%

\subsection{Experimental calibration}
\label{sec:ExperimentalCalibration}
To experimentally evaluate the performance of the optimized design, the topology optimized design given in Fig. \ref{fig:05_Design_02_05_3_rmin_2}b was fabricated and tested. The samples were 3D printed using a Connex3 Objet350 printer and FLX9785 material. Each 3D printed sample was glued to acrylic plates along the top and bottom edges in order to ensure better tester gripping. Under the uniaxial loading condition, the mechanical testing of each sample was performed on an Instron 5965 tester. Tension tests were performed at a rate of 0.1 mm/second.
\begin{figure}[!ht]
\centering
  \includegraphics[width=0.99\linewidth]{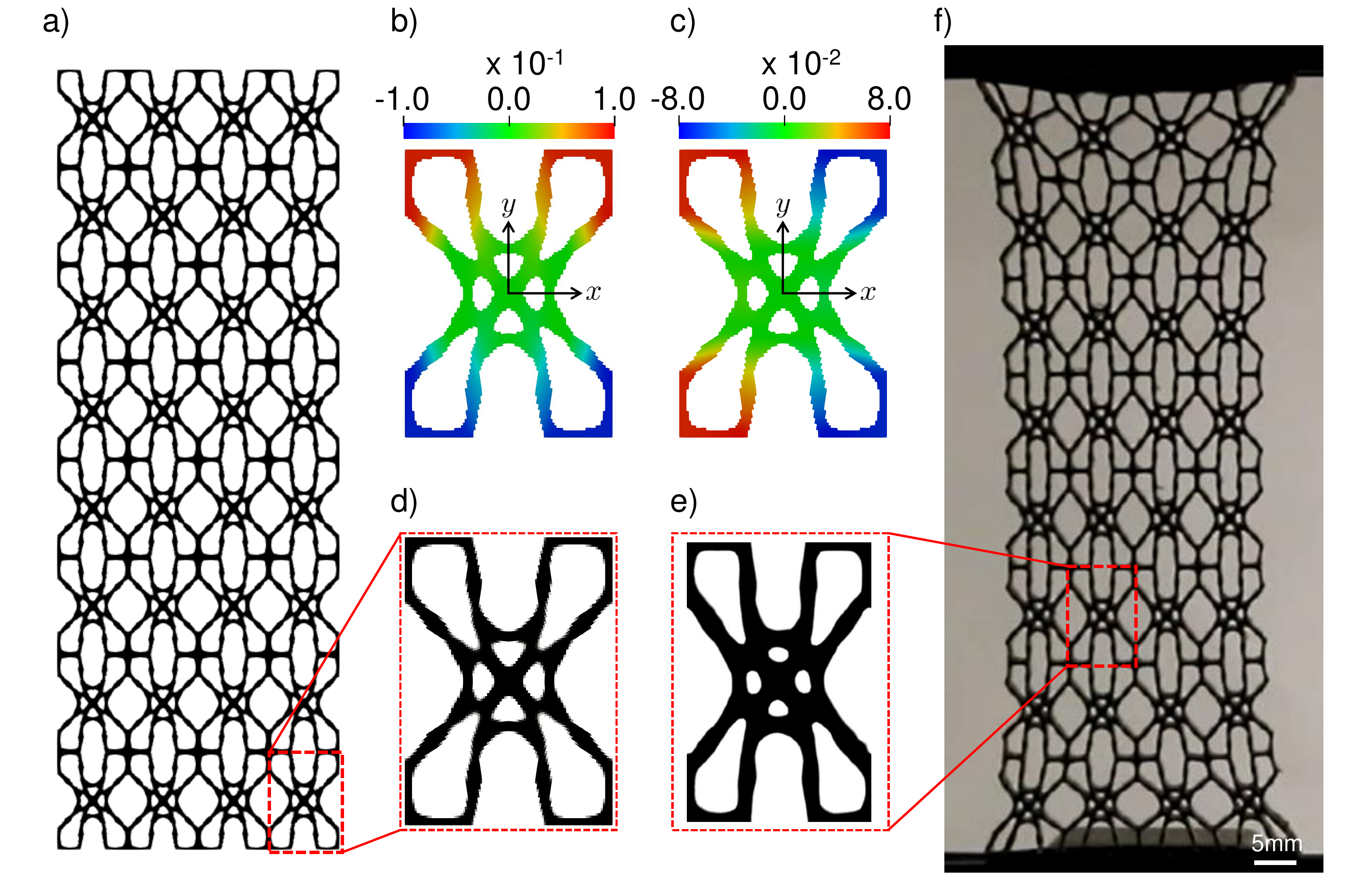}
\caption{The numerical and experimental deformed configurations of the optimized microstructure given in Fig. \ref{fig:05_Design_02_05_3_rmin_2}b at ${\widehat{E}}^{M}_{yy}$ = 20\%, a) the deformed $4 \times 8$ microstructured lattice -- simulation; b) displacement in $y$ direction -- simulation; c) displacement in $x$ direction -- simulation; d) the deformed \ac{RVE} -- simulation; e) the deformed \ac{RVE} -- experiment (background color removed thorough the image processing); and f) the deformed $4 \times 8$ microstructured lattice -- experiment.}
\label{fig:06_Design_05_03_DeformedParams}
\end{figure}

\begin{figure}[!ht]
\centering
  \includegraphics[width=0.99\linewidth]{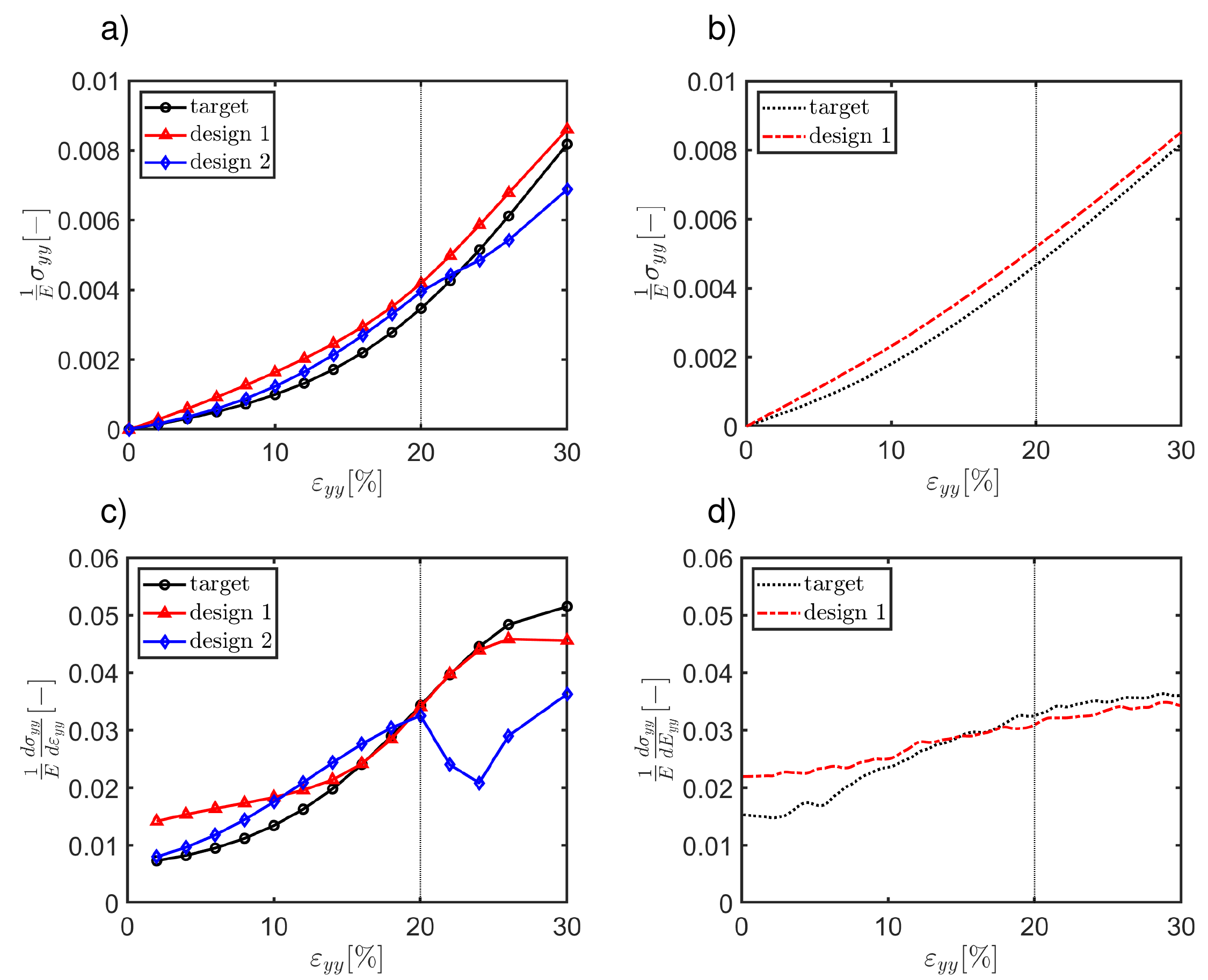}
\caption{a) The numerical stress-strain responses for the target geometry and optimized designs. b) The experimental stress-strain responses for the target geometry and design 1. c) The derivative of the numerical stress-strain responses with respect to the strain in loading direction ($\varepsilon_{yy}$). d) The derivative of the experimental stress-strain responses with respect to the strain in loading direction ($\varepsilon_{yy}$).}
\label{fig:07_Stress_Deriv_Curves_Exp_vs_Numerical}
\end{figure}

The deformed configurations of the numerical and experimental samples are shown in Fig. \ref{fig:06_Design_05_03_DeformedParams}a and Fig. \ref{fig:06_Design_05_03_DeformedParams}f. The comparison on the experimental and numerical stress-strain responses is shown in Fig. \ref{fig:07_Stress_Deriv_Curves_Exp_vs_Numerical}. The results show qualitatively similar trends in the stress-strain responses. However, the use of elastomeric type materials, and thus nonlinear elastic material response in the experimental samples could be the possible cause of the discrepancy between the experimental and numerical results. For both designs 1 and 2, it is clear that the numerical responses cross the target response at 20\% strain. For design 1, the experimental response crosses the target response close to 20\% applied strain, as intended.
%

%

%
\section{Conclusions}
\label{sec:Conclusions}
This paper presented a method for the design optimization of periodic microstructured materials with prescribed nonlinear constitutive properties over finite strain ranges. For the demonstrated examples, we considered only the geometric nonlinearity. An optimization problem was formulated to match a target mechanical response. This was accomplished by integrating a nonlinear homogenization technique into the formulation of a topology optimization algorithm that considers the effects of locally varying macroscopic strain/stress on the response of the microsctructured unit cell. Adjoint sensitivities were derived to computed the generalized sensitivities of the homogenized tangent stiffness tensor with respect to design and state variables. Two-dimensional topology optimization examples were considered to study the performance of the presented approach. An optimized design was additively manufactured and its response was calibrated through the experiment. The topology optimization approach enables design of nonlinear lattice-like materials with tailored homogenized constitutive properties, for applications ranging from impact mitigation to wearable electronics. 
%

%
\section*{Acknowledgments}
This work was supported by the Air Force Office of Scientific Research under award number FA9550-16-1-0142. R. B. thanks Dr. N. P. van Dijk for useful discussions on the homogenization technique.
%

%
\section*{References}
\bibliography{NlnrHomogenMethod_final}

\begin{thebibliography}{10}
\expandafter\ifx\csname url\endcsname\relax
  \def\url#1{\texttt{#1}}\fi
\expandafter\ifx\csname urlprefix\endcsname\relax\def\urlprefix{URL }\fi
\expandafter\ifx\csname href\endcsname\relax
  \def\href#1#2{#2} \def\path#1{#1}\fi

\bibitem{CWJ+:15}
A.~Clausen, F.~Wang, J.~S. Jensen, O.~Sigmund, J.~A. Lewis, Topology optimized
  architectures with programmable poisson's ratio over large deformations,
  Advanced Materials 27~(37) (2015) 5523--5527.

\bibitem{HLR:14}
M.~I. Hussein, M.~J. Leamy, M.~Ruzzene, Dynamics of phononic materials and
  structures: Historical origins, recent progress, and future outlook, Applied
  Mechanics Reviews 66~(4) (2014) 040802.

\bibitem{SKR+:15}
S.~Shan, S.~H. Kang, J.~R. Raney, P.~Wang, L.~Fang, F.~Candido, J.~A. Lewis,
  K.~Bertoldi, Multistable architected materials for trapping elastic strain
  energy, Advanced Materials 27~(29) (2015) 4296--4301.

\bibitem{MCJ+:16}
Q.~Ma, H.~Cheng, K.-I. Jang, H.~Luan, K.-C. Hwang, J.~A. Rogers, Y.~Huang,
  Y.~Zhang, A nonlinear mechanics model of bio-inspired hierarchical lattice
  materials consisting of horseshoe microstructures, Journal of the Mechanics
  and Physics of Solids 90 (2016) 179--202.

\bibitem{YMC+:19}
H.~Yasuda, Y.~Miyazawa, E.~G. Charalampidis, C.~Chong, P.~G. Kevrekidis,
  J.~Yang, Origami-based impact mitigation via rarefaction solitary wave
  creation, Science Advances 5~(5) (2019) eaau2835.

\bibitem{RHS:10}
J.~A. Rogers, T.~Someya, Y.~Huang, Materials and mechanics for stretchable
  electronics, Science 327~(5973) (2010) 1603--1607.

\bibitem{Nesterenko:13}
V.~Nesterenko, Dynamics of heterogeneous materials, Springer Science \&
  Business Media, 2013.

\bibitem{BK:88}
M.~P. Bends{\o}e, N.~Kikuchi, Generating optimal topologies in structural
  design using a homogenization method, Computer Methods in Applied Mechanics
  and Engineering 71~(2) (1988) 197--224.

\bibitem{BPS:00}
T.~Buhl, C.~B. Pedersen, O.~Sigmund, Stiffness design of geometrically
  nonlinear structures using topology optimization, Structural and
  Multidisciplinary Optimization 19~(2) (2000) 93--104.

\bibitem{BT:01}
T.~E. Bruns, D.~A. Tortorelli, Topology optimization of non-linear elastic
  structures and compliant mechanisms, Computer Methods in Applied Mechanics
  and Engineering 190~(26-27) (2001) 3443--3459.

\bibitem{GL:01}
H.~C. Gea, J.~Luo, Topology optimization of structures with geometrical
  nonlinearities, Computers \& Structures 79~(20-21) (2001) 1977--1985.

\bibitem{WSJ:14}
F.~Wang, O.~Sigmund, J.~S. Jensen, Design of materials with prescribed
  nonlinear properties, Journal of the Mechanics and Physics of Solids 69
  (2014) 156--174.

\bibitem{MLR:13}
K.~L. Manktelow, M.~J. Leamy, M.~Ruzzene, Topology design and optimization of
  nonlinear periodic materials, Journal of the Mechanics and Physics of Solids
  61~(12) (2013) 2433--2453.

\bibitem{XB:17}
L.~Xia, P.~Breitkopf, Recent advances on topology optimization of multiscale
  nonlinear structures, Archives of Computational Methods in Engineering 24~(2)
  (2017) 227--249.

\bibitem{GKB:10}
M.~G. Geers, V.~G. Kouznetsova, W.~Brekelmans, Multi-scale computational
  homogenization: Trends and challenges, Journal of Computational and Applied
  Mathematics 234~(7) (2010) 2175--2182.

\bibitem{GKM+:17}
M.~G. Geers, V.~G. Kouznetsova, K.~Matou{\v{s}}, J.~Yvonnet, Homogenization
  methods and multiscale modeling: Nonlinear problems, Encyclopedia of
  Computational Mechanics Second Edition (2017) 1--34.

\bibitem{NTN:13}
P.~B. Nakshatrala, D.~A. Tortorelli, K.~Nakshatrala, Nonlinear structural
  design using multiscale topology optimization. part i: Static formulation,
  Computer Methods in Applied Mechanics and Engineering 261 (2013) 167--176.

\bibitem{KWP+:19}
D.~Kumar, Z.-P. Wang, L.~H. Poh, S.~T. Quek, Isogeometric shape optimization of
  smoothed petal auxetics with prescribed nonlinear deformation, Computer
  Methods in Applied Mechanics and Engineering 356 (2019) 16--43.

\bibitem{vanDijk:16}
N.~P. van Dijk, Formulation and implementation of stress-driven and/or
  strain-driven computational homogenization for finite strain, International
  Journal for Numerical Methods in Engineering 107~(12) (2016) 1009--1028.

\bibitem{DCR+:12}
R.~De~Borst, M.~A. Crisfield, J.~J. Remmers, C.~V. Verhoosel, Nonlinear finite
  element analysis of solids and structures, John Wiley \& Sons, 2012.

\bibitem{BRG:19}
R.~Behrou, R.~Ranjan, J.~K. Guest, Adaptive topology optimization for
  incompressible laminar flow problems with mass flow constraints, Computer
  Methods in Applied Mechanics and Engineering 346 (2019) 612--641.

\bibitem{Svanberg:87}
K.~Svanberg, The method of moving asymptotes—a new method for structural
  optimization, International Journal for Numerical Methods in Engineering
  24~(2) (1987) 359--373.

\bibitem{Bendsoe:89}
M.~P. Bends{\o}e, Optimal shape design as a material distribution problem,
  Structural and Multidisciplinary Optimization 1~(4) (1989) 193--202.

\bibitem{WLS:11}
F.~Wang, B.~S. Lazarov, O.~Sigmund, On projection methods, convergence and
  robust formulations in topology optimization, Structural and
  Multidisciplinary Optimization 43~(6) (2011) 767--784.

\bibitem{KG:90}
J.~Guedes, N.~Kikuchi, Preprocessing and postprocessing for materials based on
  the homogenization method with adaptive finite element methods, Computer
  Methods in Applied Mechanics and Engineering 83~(2) (1990) 143--198.

\bibitem{KBB:01}
V.~Kouznetsova, W.~Brekelmans, F.~Baaijens, An approach to micro-macro modeling
  of heterogeneous materials, Computational Mechanics 27~(1) (2001) 37--48.

\bibitem{TN:04}
C.~Truesdell, W.~Noll, The non-linear field theories of mechanics, in: The
  Non-linear Field Theories of Mechanics, Springer, 2004, pp. 1--579.

\bibitem{LS:02}
T.-T. Lu, S.-H. Shiou, Inverses of 2$\times$ 2 block matrices, Computers \&
  Mathematics with Applications 43~(1-2) (2002) 119--129.

\bibitem{Sigmund:00}
O.~Sigmund, A new class of extremal composites, Journal of the Mechanics and
  Physics of Solids 48~(2) (2000) 397--428.

\bibitem{AGN+:10}
S.~Amstutz, S.~Giusti, A.~Novotny, E.~de~Souza~Neto, Topological derivative for
  multi-scale linear elasticity models applied to the synthesis of
  microstructures, International Journal for Numerical Methods in Engineering
  84~(6) (2010) 733--756.

\bibitem{ALS:14}
E.~Andreassen, B.~S. Lazarov, O.~Sigmund, Design of manufacturable 3d extremal
  elastic microstructure, Mechanics of Materials 69~(1) (2014) 1--10.

\bibitem{AW:05}
J.~Arora, Q.~Wang, Review of formulations for structural and mechanical system
  optimization, Structural and Multidisciplinary Optimization 30~(4) (2005)
  251--272.

\bibitem{SP:98}
O.~Sigmund, J.~Petersson, Numerical instabilities in topology optimization: a
  survey on procedures dealing with checkerboards, mesh-dependencies and local
  minima, Structural Optimization 16~(1) (1998) 68--75.

\bibitem{GPB:04}
J.~K. Guest, J.~H. Pr{\'e}vost, T.~Belytschko, Achieving minimum length scale
  in topology optimization using nodal design variables and projection
  functions, International Journal for Numerical Methods in Engineering 61~(2)
  (2004) 238--254.

\bibitem{Sigmund:07}
O.~Sigmund, Morphology-based black and white filters for topology optimization,
  Structural and Multidisciplinary Optimization 33~(4-5) (2007) 401--424.

\bibitem{SM:13}
O.~Sigmund, K.~Maute, Topology optimization approaches, Structural and
  Multidisciplinary Optimization 48~(6) (2013) 1031--1055.

\bibitem{GMS:92}
J.~R. Gilbert, C.~Moler, R.~Schreiber, Sparse matrices in matlab: Design and
  implementation, SIAM Journal on Matrix Analysis and Applications 13~(1)
  (1992) 333--356.

\bibitem{SS:86}
Y.~Saad, M.~H. Schultz, Gmres: A generalized minimal residual algorithm for
  solving nonsymmetric linear systems, SIAM Journal on Scientific and
  Statistical Computing 7~(3) (1986) 856--869.

\end{thebibliography}
%

%
\appendix
\section{Homogenization matrices}
\label{appx:HomogenMatrices}
For two spatial dimensions, the nodal coordinate matrix, $\boldsymbol{T}_{X}$, is defined as follows:
\begin{equation}\label{eq:PeriodicBCs_Tx}
\begin{split}
\boldsymbol{T}_{X}
=
\begin{bmatrix}
\vdots & \vdots & \vdots & \vdots & \vdots & \vdots \\ 
 X^{i}_{1} & X^{i}_{2} & 0 & 0 & 0 & 0 \\ 
 0 & 0 & X^{i}_{1} & X^{i}_{2} & 0 & 0 \\ 
 0 & 0 & 0 & 0 & X^{i}_{1} & X^{i}_{2} \\ 
\vdots & \vdots & \vdots & \vdots & \vdots & \vdots
\end{bmatrix}_{2n \times 4}.
\end{split}
\end{equation}
The following relationship is held between the symmetric displacement gradient vector, $\boldsymbol{\widehat{G}}^{M}_{\mathrm{sym}}$, the original displacement gradient, $\boldsymbol{\widehat{G}}^{M}$, and the Green-Lagrange strain tensors (under the assumption of the rotation-free deformation) \cite{vanDijk:16}:
\begin{equation}\label{eq:MacroMicroScaleStressStrainMeasures_04}
\begin{split}
\boldsymbol{\widehat{G}}^{M} &= \boldsymbol{T}_{s} \boldsymbol{\widehat{G}}^{M}_{\mathrm{sym}}, \\
\boldsymbol{\widehat{E}}^{M} &= \Big( \boldsymbol{\bar I} + \frac{1}{2} \boldsymbol{\bar G}^{M} \Big) \boldsymbol{\widehat{G}}^{M}_{\mathrm{sym}},
\end{split}
\end{equation}
where
\begin{equation}\label{eq:MacroMicroScaleStressStrainMeasures_05}
\begin{split}
\boldsymbol{T}_{s} 
=
\begin{bmatrix}
1 & 0 & 0 & 0 & 0 & 0 & 0 & 0 & 0\\ 
0 & 0 & 0 & 0 & 1 & 0 & 0 & 0 & 0\\ 
0 & 0 & 0 & 0 & 0 & 0 & 0 & 0 & 1\\ 
0 & 1 & 0 & 1 & 0 & 0 & 0 & 0 & 0\\ 
0 & 0 & 0 & 0 & 0 & 1 & 0 & 1 & 0\\ 
0 & 0 & 1 & 0 & 0 & 0 & 1 & 0 & 0
\end{bmatrix}
^{T},
\end{split}
\end{equation}
The matrix $\boldsymbol{\bar G}^{M}$ collects the components of the symmetric displacement gradient, $\boldsymbol{\widehat{G}}^{M}_{\mathrm{sym}}$, is given as follows:
\begin{equation}\label{eq:MacroMicroScaleStressStrainMeasures_06}
\begin{split}
\boldsymbol{\bar G}^{M}
=
\begin{bmatrix}
G^{M}_{11} & 0 & 0 & G^{M}_{12} & 0 & G^{M}_{31} \\ 
0 & G^{M}_{22} & 0 & G^{M}_{12} & G^{M}_{23} & 0 \\ 
0 & 0 & G^{M}_{33} & 0 & G^{M}_{23} & G^{M}_{31} \\ 
G^{M}_{12} & G^{M}_{12} & 0 & (G^{M}_{11} + G^{M}_{22}) & G^{M}_{31} & G^{M}_{23} \\ 
0 & G^{M}_{23} & G^{M}_{23} & G^{M}_{31} & (G^{M}_{22} + G^{M}_{33}) & G^{M}_{12} \\ 
G^{M}_{31} & 0 & G^{M}_{31} & G^{M}_{23} & G^{M}_{12} & (G^{M}_{33}+G^{M}_{11})
\end{bmatrix}.
\end{split}
\end{equation}
The diagonal matrix, $\boldsymbol{\bar I}$, is defined as follows:
\begin{equation}\label{eq:MacroMicroScaleStressStrainMeasures_07}
\begin{split}
\boldsymbol{\bar I}
=
\begin{bmatrix}
1 &  &  &  &  & \\ 
 & 1 &  &  & \mathbf{0} & \\ 
 &  & 1 &  &  & \\ 
 &  &  & 2 &  & \\ 
 & \mathbf{0} &  &  & 2 & \\ 
 &  &  &  &  & 2
\end{bmatrix}.
\end{split}
\end{equation}
The so-called geometrical stress, $\boldsymbol{\bar S}^{M}$, is be derived from the components of the internal macroscopic second Piola-Kirchhoff stress, $\widehat{\boldsymbol{S}}^{M}_{\mathrm{int}}$:
\begin{equation}\label{eq:LagrangeMultiplierApproach_08}
\begin{split}
\boldsymbol{\bar S}^{M}
=
\begin{bmatrix}
S_{11} & 0 & 0 & S_{12} & 0 & S_{31} \\ 
0 &S_{22} & 0 & S_{12} & S_{23} & 0 \\ 
0 & 0 & S_{33} & 0 & S_{23} & S_{31} \\ 
S_{12} & S_{12} & 0 & (S_{11} + S_{22}) & S_{31} & S_{23} \\ 
0 & S_{23} & S_{23} & S_{31} & (S_{22} + S_{33}) & S_{12} \\ 
S_{31} & 0 & S_{31} & S_{23} & S_{12} & (S_{33}+S_{11})
\end{bmatrix}.
\end{split}
\end{equation}
\section{Microscale internal energy and its first and second derivatives}
\label{appx:DerivativeofStiffnessMatrix}
At the microscale and in the absence of external forces, the variational form of the internal strain energy can be written as:
\begin{equation}\label{eq:MicroEquilibriumEq_01}
\begin{split}
\delta \mathcal{W}^{\mu}_{\mathrm{int}} (\boldsymbol{\widehat{u}}) = \int_{\Omega_{\mathrm{rve}}} S_{IJ}^{\mu} (E_{IJ}^{\mu} (\boldsymbol{\widehat{u}}) ) \ \delta E_{IJ}^{\mu} (\boldsymbol{\widehat{u}}) \ d\Omega,
\end{split}
\end{equation}
where $\boldsymbol{E}^{\mu}$ and $\boldsymbol{S}^{\mu}$ is the Green-Lagrange strain and the Second Piola-Kirchhoff stress tensor, respectively. Linearization of the in internal strain energy leads to the following derivatives:
\begin{equation}\label{eq:MicroEquilibriumEq_02}
\begin{split}
\frac{\partial \delta \mathcal{W}^{\mu}_{\mathrm{int}}}{\partial \widehat{u}_{m}} 
&= \int_{\Omega_{\mathrm{rve}}} \Big( \frac{\partial \delta E_{IJ}^{\mu}}{\partial \widehat{u}_{m}} S_{IJ}^{\mu} + \delta E_{IJ}^{\mu} C_{IJKL}^{\mu} \frac{\partial E_{KL}^{\mu}}{\partial \widehat{u}_{m}} \Big) \ d\Omega.
\end{split}
\end{equation}
And the second derivatives gives:
\begin{equation}\label{eq:MicroEquilibriumEq_021}
\begin{split}
\frac{\partial}{\partial \widehat{u}_{t}} \Big( \frac{\partial \delta \mathcal{W}^{\mu}_{\mathrm{int}}}{\partial \widehat{u}_{m}} \Big) 
&= \int_{\Omega_{\mathrm{rve}}} \Big( \frac{\partial \delta E_{IJ}^{\mu}}{\partial \widehat{u}_{m}} C_{IJKL}^{\mu} \frac{\partial E_{KL}^{\mu}}{\partial \widehat{u}_{t}} + \frac{\partial \delta E_{IJ}^{\mu}}{\partial \widehat{u}_{t}} C_{IJKL}^{\mu} \frac{\partial E_{KL}^{\mu}}{\partial \widehat{u}_{m}} \\
&+ \delta E_{IJ}^{\mu} \frac{\partial C_{IJKL}^{\mu}}{\partial \widehat{u}_{t}} \frac{\partial E_{KL}^{\mu}}{\partial \widehat{u}_{m}} + \delta E_{IJ}^{\mu} C_{IJKL}^{\mu} \frac{\partial}{\partial \widehat{u}_{t}} \Big( \frac{\partial E_{KL}^{\mu}}{\partial \widehat{u}_{m}} \Big) \Big) \ d\Omega.
\end{split}
\end{equation}
The microscale Green-Lagrange strain tensor and it's derivatives is expressed as:
\begin{equation}\label{eq:MicroEquilibriumEq_03}
\begin{split}
E_{KL}^{\mu} = \frac{1}{2} \big( (F_{Kp}^{\mu})^{T} F_{pL}^{\mu} - \delta_{KL} \big), \qquad F_{pL}^{\mu} = \frac{\partial u_{p}}{\partial X_{L}^{\mu}} + \delta_{pL},
\end{split}
\end{equation}
\begin{equation}\label{eq:MicroEquilibriumEq_04}
\begin{split}
\frac{\partial E_{KL}^{\mu}}{\partial \widehat{u}_{m}} = \frac{1}{2} \Big( \frac{\partial (F_{Kp}^{\mu})^{T}}{\partial \widehat{u}_{m}} F_{pL}^{\mu} + (F_{Kp}^{\mu})^{T} \frac{\partial F_{pL}^{\mu}}{\partial \widehat{u}_{m}} \Big),
\end{split}
\end{equation}
\begin{equation}\label{eq:MicroEquilibriumEq_05}
\begin{split}
\delta E_{IJ}^{\mu} = \frac{1}{2} \Big( F_{iJ}^{\mu} \frac{\partial \delta u_{i}}{\partial X_{I}^{\mu}} + F_{iI}^{\mu} \frac{\partial \delta u_{i}}{\partial X_{J}^{\mu}} \Big),
\end{split}
\end{equation}
\begin{equation}\label{eq:MicroEquilibriumEq_06}
\begin{split}
\frac{\partial \delta E_{IJ}^{\mu}}{\partial \widehat{u}_{m}} = \frac{1}{2} \Big( \frac{\partial F_{iJ}^{\mu}}{\partial \widehat{u}_{m}} \frac{\partial \delta u_{i}}{\partial X_{I}^{\mu}} + \frac{\partial F_{iI}^{\mu}}{\partial \widehat{u}_{m}} \frac{\partial \delta u_{i}}{\partial X_{J}^{\mu}} \Big),
\end{split}
\end{equation}
\begin{equation}\label{eq:MicroEquilibriumEq_07}
\begin{split}
\frac{\partial}{\partial \widehat{u}_{t}} \Big( \frac{\partial E_{KL}^{\mu}}{\partial \widehat{u}_{m}} \Big) = \frac{1}{2} \Big( \frac{\partial (F_{Kp}^{\mu})^{T}}{\partial \widehat{u}_{m}} \frac{\partial F_{pL}^{\mu}}{\partial \widehat{u}_{t}} + \frac{\partial (F_{Kp}^{\mu})^{T}}{\partial \widehat{u}_{t}} \frac{\partial F_{pL}^{\mu}}{\partial \widehat{u}_{m}} \Big).
\end{split}
\end{equation}
The material tangent stiffness matrix, $C_{IJKL}^{\mu}$, can be computed from the given microscopic constitutive law. The vectorized forms of Eq.~\eqref{eq:MicroEquilibriumEq_01} is associated with the internal force vector, $\boldsymbol{f}_{\mathrm{int}}^{\mu} (\boldsymbol{\widehat{u}})$. Eqs.~\eqref{eq:MicroEquilibriumEq_02} and \eqref{eq:MicroEquilibriumEq_021} result in the microscale tangent stiffness matrix and its derivatives.

\end{document}